\documentclass[a4paper,11pt]{article}
\pdfoutput=1
\usepackage{jinstpub}
\usepackage{graphicx}
\usepackage{caption}
\usepackage{subcaption}
\usepackage{lineno}
\usepackage[font=scriptsize,labelfont=bf,skip=6pt]{caption}
\pdfoutput=1 
\usepackage{multirow}
\usepackage[some]{background}
\SetBgScale{1}
\SetBgContents{\parbox{10cm}{%
  \Huge Accepted in JINST\_082P\_1221:  \today\\[14cm]\rotatebox{180}{\Huge Draft:  \today}}}
\SetBgColor{gray}
\SetBgAngle{270}
\SetBgOpacity{0.2}

\title{\boldmath The Qualification of GEM detector and its application to Imaging}

\author[a,1]{A.~Ahmed, \note{Corresponding author.}}
\author[a]{A.~Kumar,}
\author[a]{Md. Naimuddin,}
\author[b,c]{M. Babij,}
\author[b,c]{and~P. Bielowka}

\affiliation[a]{Department of Physics $\&$ Astrophysics\\ University of Delhi, Delhi, India}
\affiliation[b]{TTA Techtra Ltd, Wroclaw, Poland}
\affiliation[c]{Department of Microsystems, Faculty of Electronics, Photonics and Microsystems \\Wroclaw University of Science and Technology, Poland.}
\emailAdd{asar0786@gmail.com}

\begin{document}
\BgThispage
\maketitle
\begin{abstract}
~The Gas Electron Multiplier (GEM) is a new age detector, which can handle  the high flux of particles. The GEM foil, which is constructed using 50$\mu$m highly insulating foil (Kapton/Apical) coated with 5$\mu$m layers of copper, on both sides, with a network of specifically shaped holes is the major component of these detectors. The European Center for Nuclear Research (CERN) has been the sole supplier of the GEM foils until recently when a few other companies started manufacturing GEM foils under the transfer of technology (TOT) agrement from CERN. Techtra is one such company in Europe which gained a right to use CERN developed technology in order to produce commercially viable GEM foils. Micropack Pvt. Ltd. is another company in India which has successfully manufactured good quality GEM foils. Due to the microscopic structure of holes and dependence on the electric field inside, it becomes essential to study the defect and uniformity of holes along with the electrical property of foils under ambient conditions. In this work we are reporting the tests condition of Techtra GEM foils. We report on the development of a cost effective and efficient technique to study the GEM foils holes geometry, distribution, and defects. We also report on the electrical properties of these foils like leakage current, stability, and discharges. At the detector level, we describe the high voltage (HV) response, gain, uniformity, and stability. The GEMs have been proposed to have a wider applications, so we performed a feasibility study to utilize these for the imaging. We irrediated various objects of varying density with X-rays and reconstructed the images. The reconstructed image shows a good distinction between materials of different densities, which can be very useful in various applications like medical imaging or cargo imaging.
\end{abstract}

\begin{keywords}
~Keywords: GEM, Characterization, Stability, Uniformity, Imaging, Techtra
\end{keywords}
\section{Introduction}
\label{sec:intro}
F. Sauli in 1997 \cite{one} gave the concept of GEM to overcome the high flux rate handling and spatial resolution in other gaseous detectors. GEM foils consist of a 50 $\mu$m thin Kapton foil coated with 5$\mu$m thin layer of copper on both sides. Bi-conical holes with 40-60 $\mu$m inner and 60-80 $\mu$m outer diameters are chemically etched in the foil at a pitch of about 140 $\mu$m by using either a double mask or single mask \cite{one_01} technique.
\begin{figure}[!ht]
    \centering
    \begin{subfigure}[b]{0.46\textwidth}
		\includegraphics[width=6cm, height=4cm]{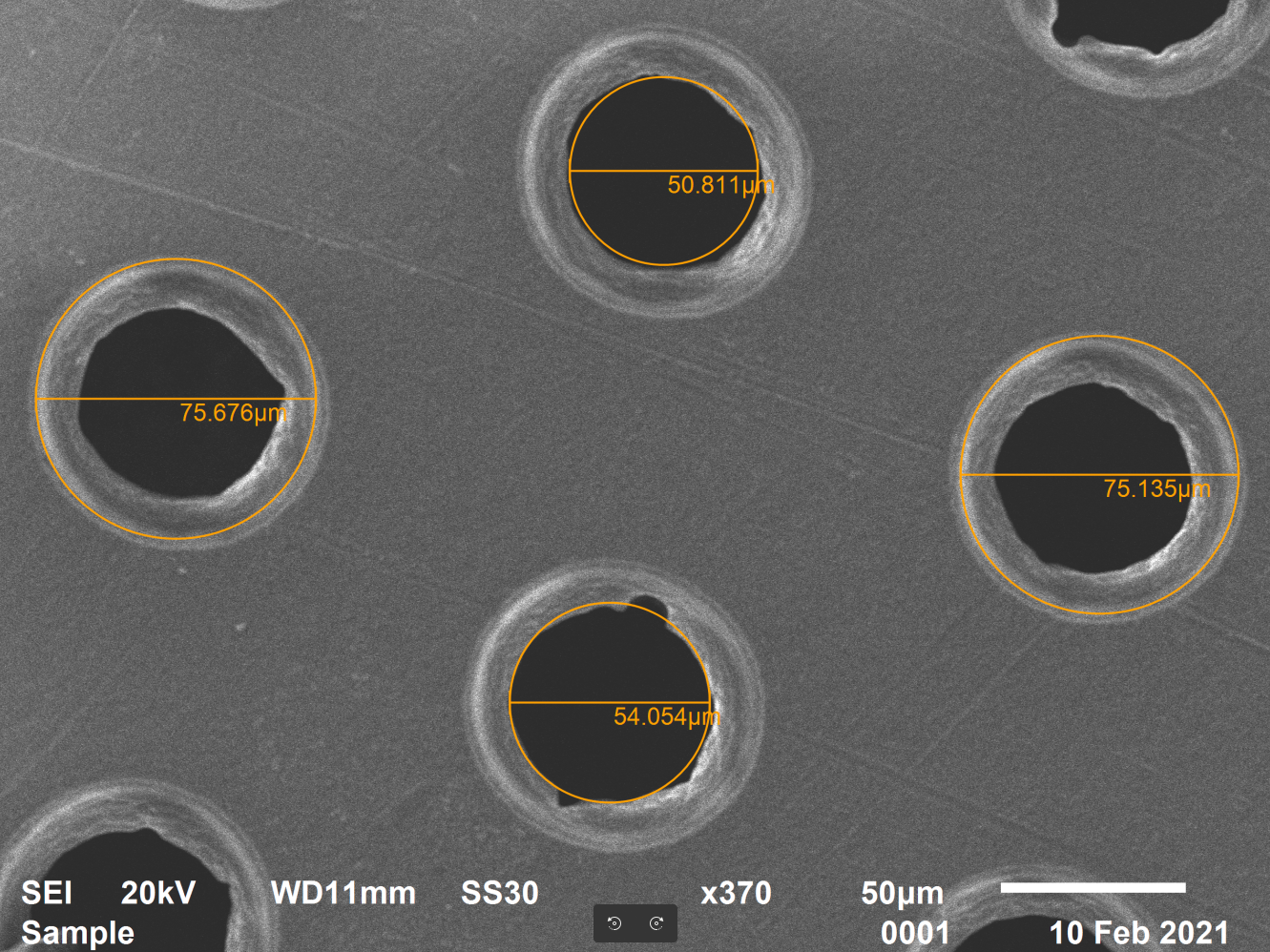}
\qquad
        \caption{ }
    \end{subfigure}
    \begin{subfigure}[b]{0.46\textwidth}
        \includegraphics[width=6cm, height=4.0 cm]{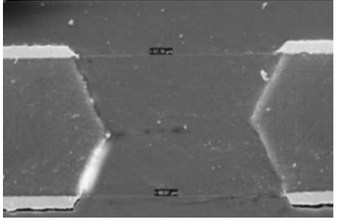}
        \caption{ }
    \end{subfigure}
   \caption{(a) Top view of the GEM foil using SEM, (b) Cross-sectional view of the foil showing the double cone structure of the engraved holes. } \label{fig:Foil_and_Cone}
\end{figure}
GEM foils have attracted significant interest from the nuclear and particle physics communities, as they are excellent candidates to be used in a high flux environment as a tracking detectors. This detector technology has been used successfully as a tracking detector in many experiments, such as STAR \cite{two}, LHCb \cite{lhcb}, TOTEM \cite{two_02}, COMPASS \cite{compass}, CMS \cite{six_03} and ALICE \cite{four}, and is expected to be used in many future experiments and their upgrades. 
Untill now, CERN was the main producer of small as well as large area GEM foils. Various companies like Micropack (India) \cite{six}, Techtra (Poland) \cite{techtra}, Mecaro (South Korea) \cite{mecaro}, etc, have also succeeded in producing GEM foils 
with both double-mask
and single-mask
technology.
Generally, foil production begins with substrate preparation (5 $\mu$m copper is coated on either side of 50 $\mu$m Apical Type NP/Kapton film), photo-resistive coating (15 $\mu$m thick photo-resistive layer is applied on both sides of the substrate), masking (a hole patterned mask is placed on the top or either side of the substrate), UV- light exposure to transfer hole pattern to the substrate and several solvents and acid bath ends with the final GEM holes pattern in the foil. Double mask technology has the limitation of mask alignment problem for larger foils but has good control over small size foils, while for larger foils a minor misalignment of the mask can propagate the error to the whole foil. The single-mask technique overcomes the rigorous practice of alignment of two masks and allows the production of foils with a large area. However, the holes obtained with refinements in the single-mask technique are asymmetrically bi-conical \cite{ashaq_single_mask} in shape, discussed in Section ~\ref{Optical_Assessment}, compared to symmetrically bi-conical holes from double-mask technology.
In order to qualify these GEM foils as commercially and scientifically reliable, several quality control tests needs to be performed. In this paper, we will describe the technique used for foil qualification, details of the quality control tests, and various characterization studies performed on three Techtra GEM foils for them to be used in GEM detectors for various applications such as imaging or scanning for commercial purposes.\\
\section{Optical Assessment}
\label{Optical_Assessment}
\begin{figure}[!ht]
    \centering
        \includegraphics[width=10cm, height=7cm]{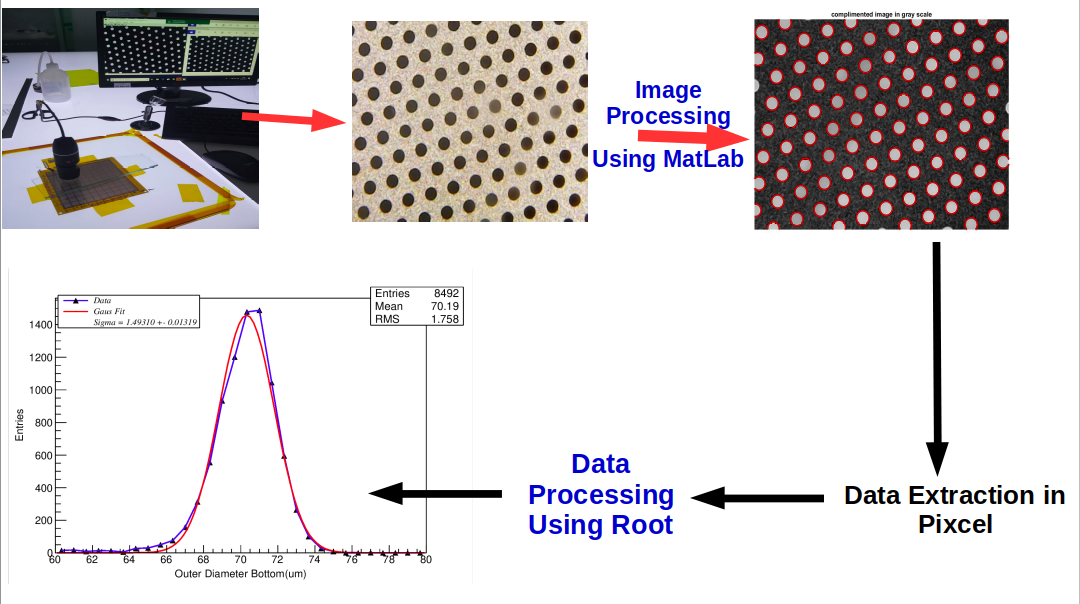}
   \caption{Flow diagram of the setup used for the optical measurements.}   \label{fig:Optical_Sketch}
\end{figure}

The quality of GEM foil, i.e., hole geometry \cite{six_01}, pattern, and flaws, has a significant impact on the performance of GEM detectors. The charge multiplication and charging up phenomenon is influenced considerably more by hole size and pitch \cite{pitch_performance}. There are roughly 600k holes in a 10 cm $\times$ 10 cm GEM foil with a surface hole diameter of 70 $\mu$m and a pitch of 140 $\mu$m. Any asymmetry or faults in the hole pattern and its shape can have a direct impact on the foil's performance, and thus the detector's performance. As a result, it's critical to investigate the geometry of holes, imperfections, and contaminants on foils that can lead to foil failure. Although there are many pricey optical microscopes available to see through these minuscule holes, we focused on designing a cost-effective and flexible optical study system. Figure~\ref{fig:Optical_Sketch} depicts the flow diagram for the optical measurement setup. 

%

\begin{figure}[!ht]
    \begin{subfigure}[b]{0.45\textwidth}
        \includegraphics[width=6.0cm, height=5.0cm]{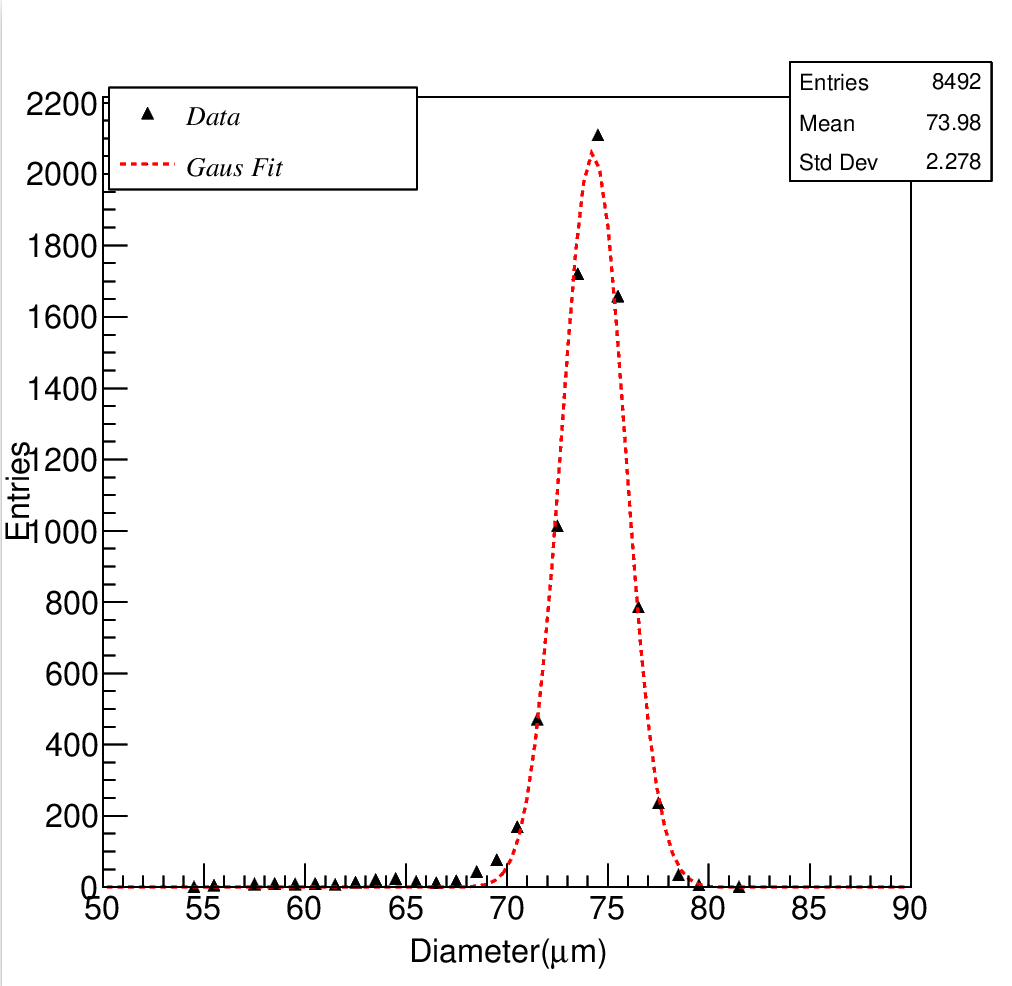}
        \caption{ }
        \label{fig:dia_inner}
    \end{subfigure}
    \begin{subfigure}[b]{0.45\textwidth}
        \includegraphics[width=6.0cm, height=5.0cm]{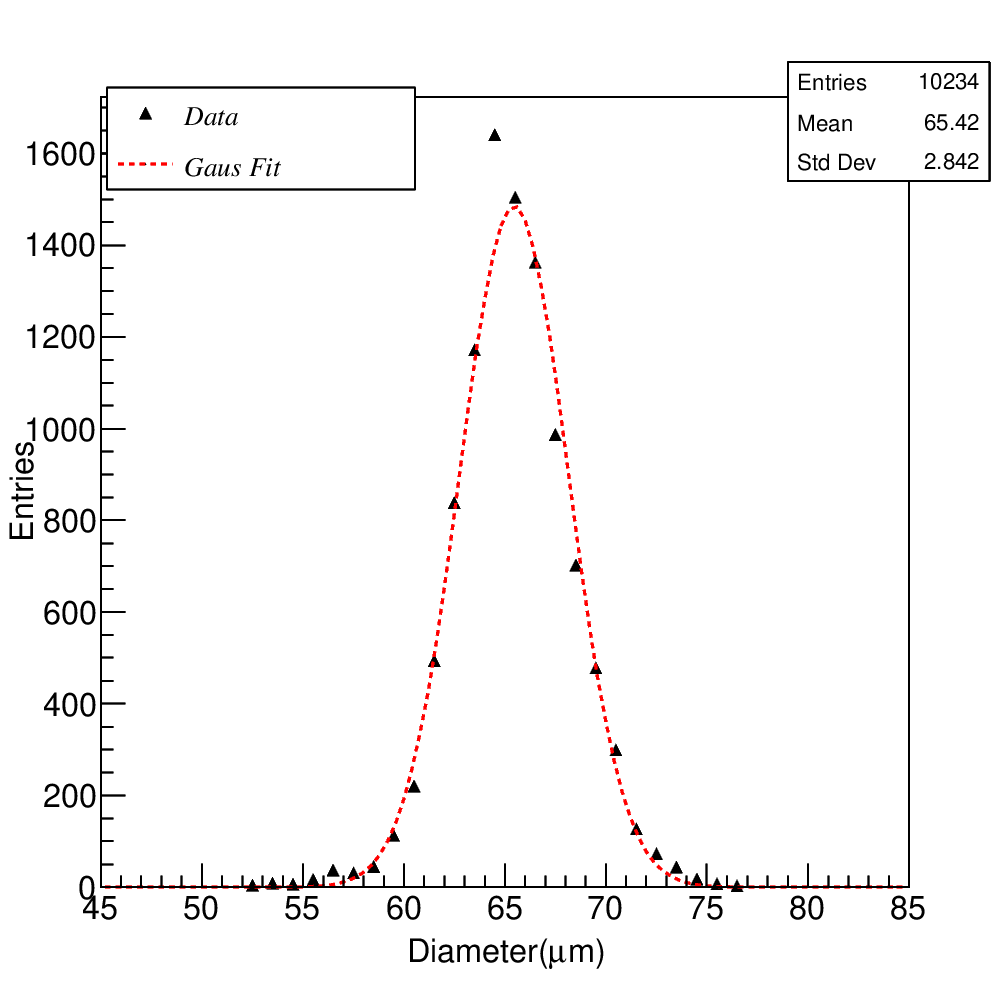}
        \caption{ }
        \label{fig:dia_outer}
    \end{subfigure}
   \caption{Outer hole diameter distribution in copper (a) top, (b) bottom.} \label{fig:Dia_copp}
\end{figure}

\begin{figure}[!ht]
    \begin{subfigure}[b]{0.45\textwidth}
        \includegraphics[width=6.0cm, height=5.0cm]{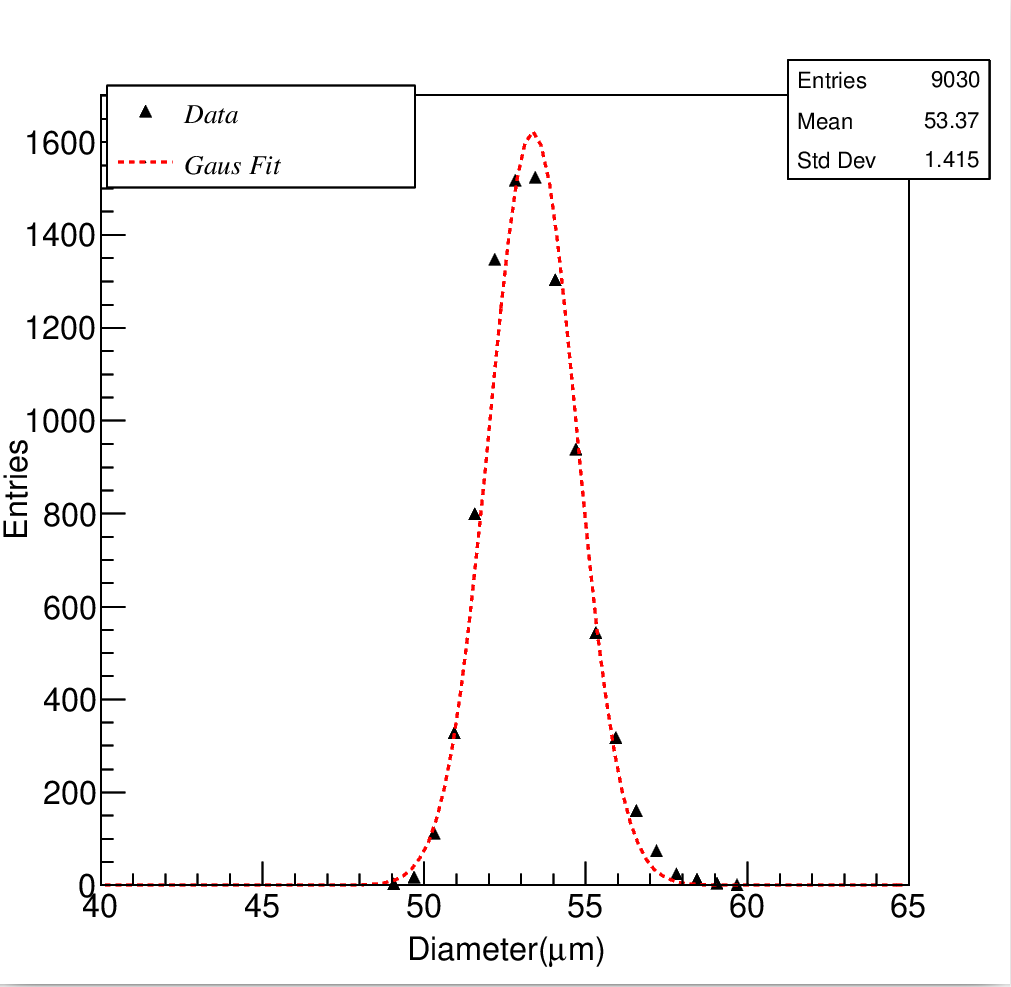}
        \caption{ }
        \label{fig:dia_inner}
    \end{subfigure}
    \begin{subfigure}[b]{0.45\textwidth}
        \includegraphics[width=6.0cm, height=5.0cm]{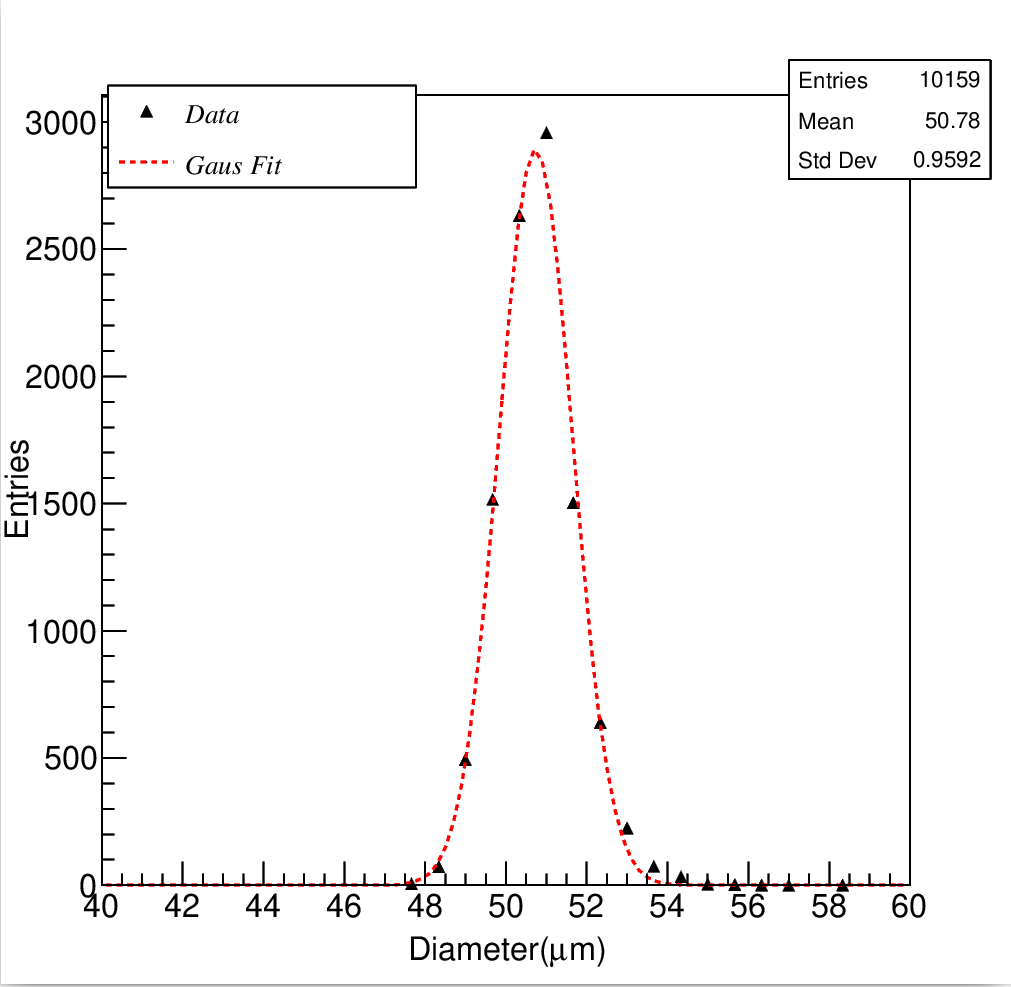}
        \caption{ }
        \label{fig:dia_outer}
    \end{subfigure}
   \caption{Inner hole diameter distribution in Kapton (a) top, (b) bottom.} \label{fig:Dia_kapton}
\end{figure}

We used a 10 cm $\times$ 10 cm GEM foil and protected it from exterior interactions by sandwiching it between two 2 mm glass plates. To take photos, a 100-cell array of 1 cm $\times$ 1 cm was drawn on a glass plate. Each cell was imaged using a digital microscope with magnification ranging from 40X to 1000X. Both sides of the GEM foil were photographed with correct adjustments. To take photos of inner hole geometry, the foil is illuminated from the reverse. All of the photos were then analysed in MATLAB \cite{O_one}, using an algorithm to locate circles and their centres in 2D, as well as determine the diameter and pitch in pixels for each hole in the image. Specifications from SEM pictures were used to calibrate the pixel data and ROOT \cite{root} was used to create histograms using data from all 100 cells in pixels and error of one standard deviation is used to quote the result. As illustrated in Figures \ref{fig:Dia_copp} $\&$ \ref{fig:Dia_kapton}, the top and bottom hole mean diameters observed on copper estimated using Gaussian fit are 73.98 $\pm$ 2.28 $\mu$m and 65.42 $\pm$ 2.84 $\mu$m respectively, whereas the inner diameter in Kapton is 53.37 $\pm$ 1.42 $\mu$m and 50.78 $\pm$ 0.96 $\mu$m. The optical measurement yielded a pitch of 140.0 $\pm$ 1.4 $\mu$m. These measurements reveal the asymmetrical nature of the bi-conical hole, which is caused by the single mask approach, in which the top layer serves as a mask for the bottom layer and is subjected to chemical etching for a longer period of time than the bottom layer.  


\section{Electrical Assessment}
\label{electrical}
The optical test reveals the hole geometry, pitch, and imperfections in the foils, whereas the electrical test reveals leakage current, cleanliness, and the capacity to tolerate high voltage without sparking.
The foils were tested in a cleanroom of class 100 \cite{cleanroom} for a rapid inspection after being properly cleaned with an adhesive roller and an insulation tester MIT Megger 420 \cite{megger}, which checks conductivity/impedance. At a potential difference of 550V, room temperature of 21$^{\circ}$C, and relative humidity of 50$\%$, all three foils required for detector assembly showed impedance of more than 100 G$\Omega$ and leakage current less than 1 nA. The current should not fluctuate over time in order for the GEM foils to perform better, as this will impact the potential difference across the foil, and hence the electric field inside the holes. This can only be established by continuously measuring its leakage current in a well-controlled environment for a prolonged period of time. For this test, foils were put in an isolated enclosure with nitrogen flushing to manage humidity and temperature, as well as a temperature and humidity monitoring system. A KANOMAX \cite{fourteen} dust particle counter Model 3887 was used to keep track of the particle count. Keithley 6517B \cite{picoamp} multimeter was utilised for higher precision in current measuring. Keithley software was used to record the measurements, which consisted of a bare GEM foil driven by a Keithley 6517B electrometer connected to the computer through a GPIB interface.
With an accuracy of $\pm$0.2 $\%$	, the current measurement range was set from 0 to 20 nA.

\begin{figure}[!ht]
    \centering
    \begin{subfigure}[b]{0.5\textwidth}
        \includegraphics[width=6.0cm, height=5.0cm]{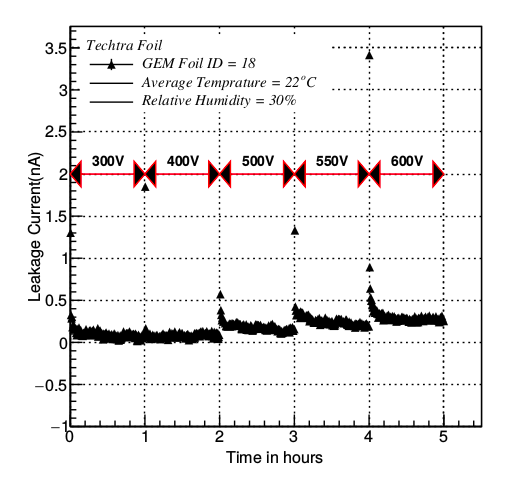}
        \caption{ }
        \label{fig:GEM1}
    \end{subfigure}
    \begin{subfigure}[b]{0.46\textwidth}
        \includegraphics[width=6cm, height=5cm]{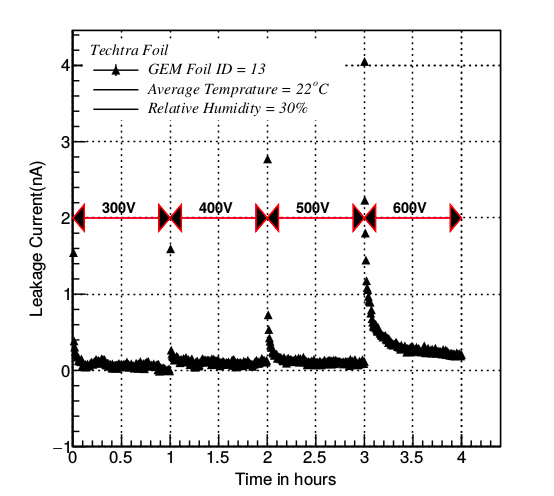} 
        \caption{ }
        \label{fig:GEM2}
    \end{subfigure}
    \vskip\baselineskip
    \begin{subfigure}[b]{0.5\textwidth}
        \includegraphics[width=6cm, height=5cm]{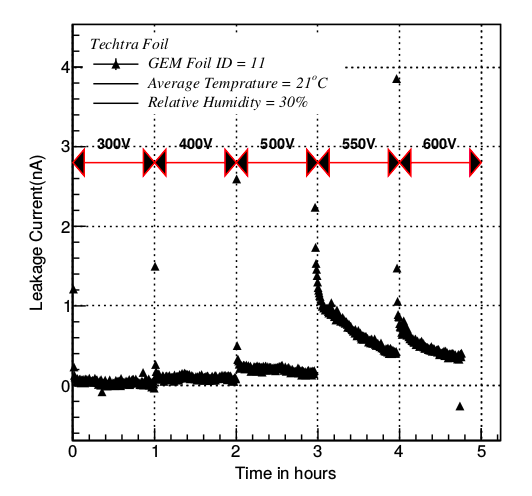}
        \caption{ }
        \label{fig:GEM3}
    \end{subfigure}
    \begin{subfigure}[b]{0.46\textwidth}
        \includegraphics[width=6cm, height=5cm]{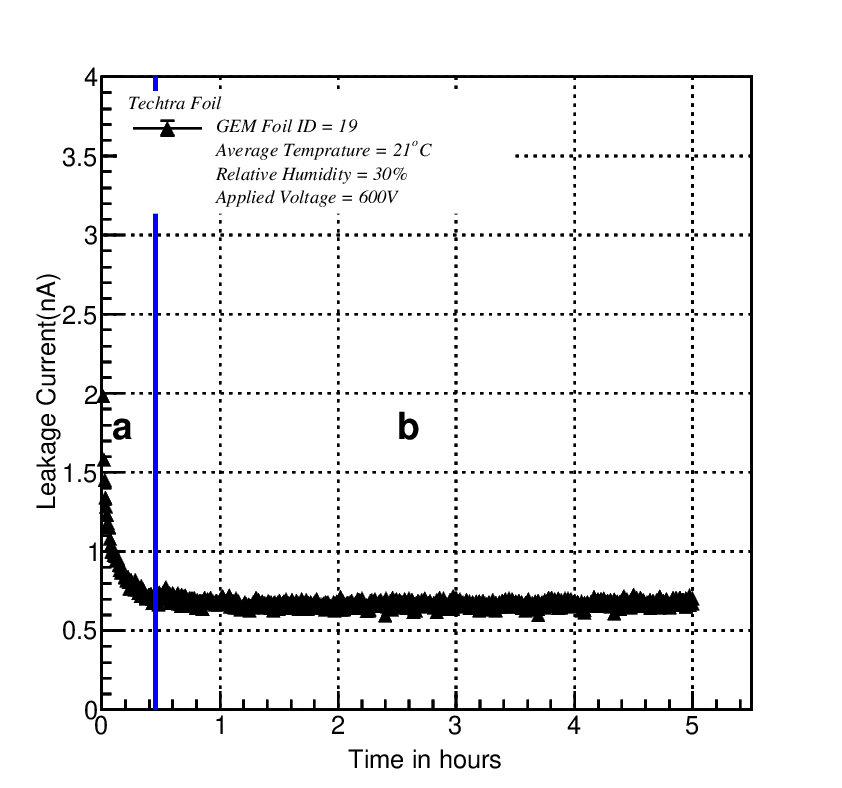} 
        \caption{ }
        \label{fig:5h}
    \end{subfigure}
   \caption{Leakage Current of Techtra GEM (a) First foil, (b) second foil, (c) third foil, (d) longer duration test at 600V, at average temperature of T=21$^{\circ}$C and relative humidity of 30\%.} \label{fig:Leakage_current}
\end{figure}

Foils were tested at a temperature of 21$^{\circ}$C and a relative humidity of 30$\%$ while being applied a voltage of 300 V to 600 V in 100 V/h steps. As demonstrated in Figure \ref{fig:Leakage_current}, all three foils performed similarly, with leakage currents under 1 nA and no visible discharges. One of the foils was tested for a total of 5 hours at 600V under the same conditions for the longer duration test. Figure \ref{fig:5h} demonstrates that leakage current reduces from 1.983 nA to 0.713 nA ($\sim64\%$ change) in the first half hour, then decreases to 0.694 nA ($\sim3\%$ change) in the second half hour, and then remains constant for the remaining  measurement time. CERN and Micropack foils \cite{ashaq} had similar results. 

\begin{figure}[ht]
    \centering
    \begin{subfigure}[b]{0.53\textwidth}
        \includegraphics[width=6cm, height=5cm]{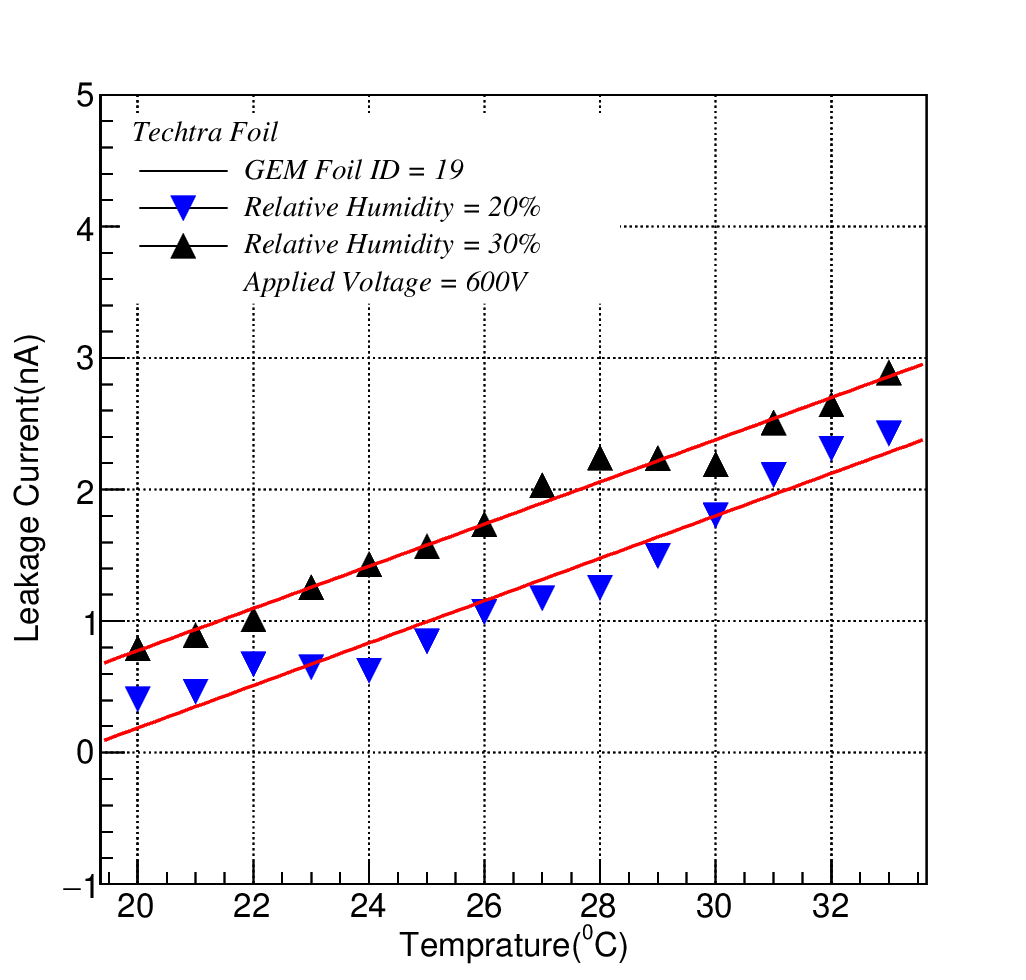}
        \caption{ }
        \label{fig:Temp}
    \end{subfigure}
    \begin{subfigure}[b]{0.46\textwidth}
        \includegraphics[width=6cm, height=5cm]{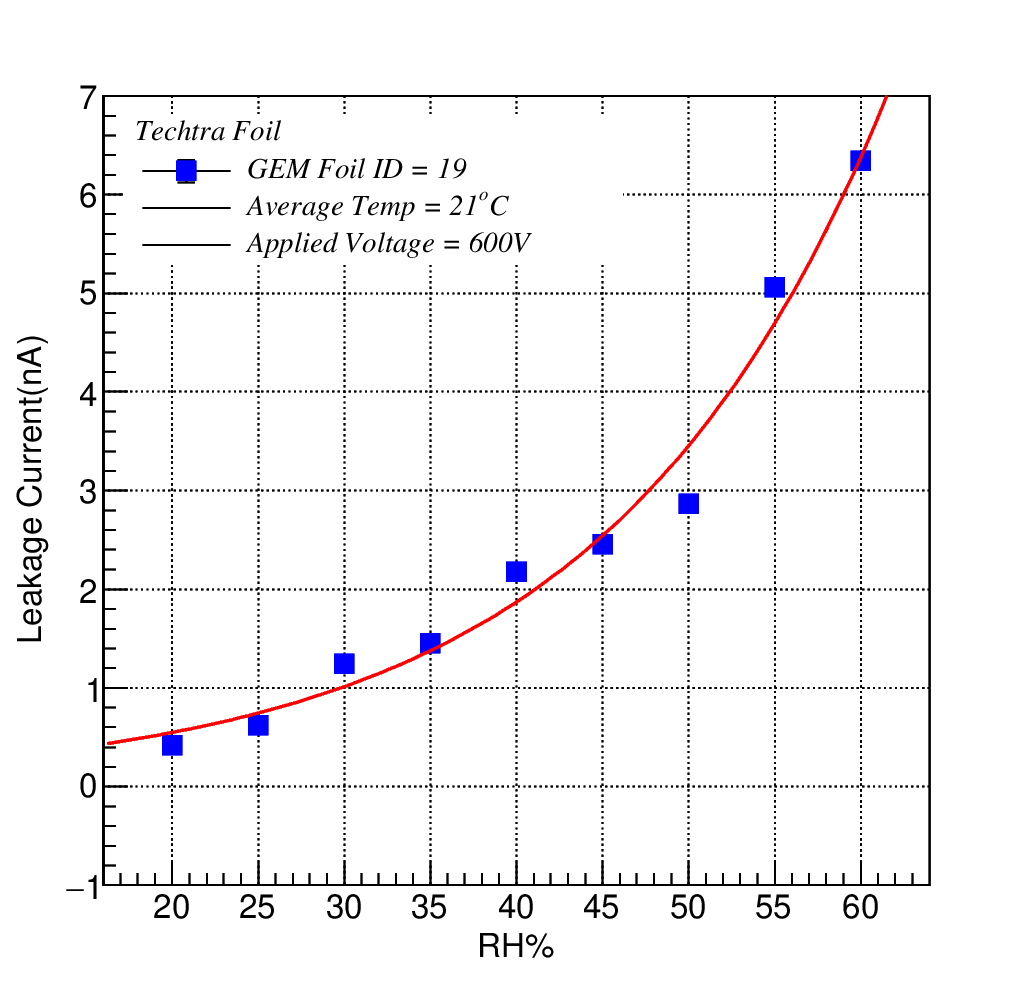}
        \caption{ }
        \label{fig:RH}
    \end{subfigure}
   \caption{Leakage current in dry nitrogen environment at 600V: (a) Varying temprature and fixed RH. (b) Varying RH and fixed temprature. The  fit shows the linear behavior of leakage current with temperature while exponential behavior with RH.} \label{fig:Temp_RH}
\end{figure}

At a fixed high voltage of 600 V, more tests were conducted to see how leakage current varied with temperature and humidity. Two sets of data were gathered, one with altering temperature and the other with varying humidity while all other parameters were held constant. Figure \ref{fig:Temp_RH} illustrates how GEM foils are extremely sensitive to humidity and temperature, causing leakage current to grow linearly with temperature and exponentially with relative humidity (RH). Leakage current should be less than 1 $n$A, which translates to a temperature of roughly 20$^\circ$C  and a relative humidity of around 25$\%$ for better performance and stability of the detector. 


\begin{figure}[!ht]
    \centering
    \begin{subfigure}[b]{0.53\textwidth}
        \includegraphics[width=6cm, height=5cm]{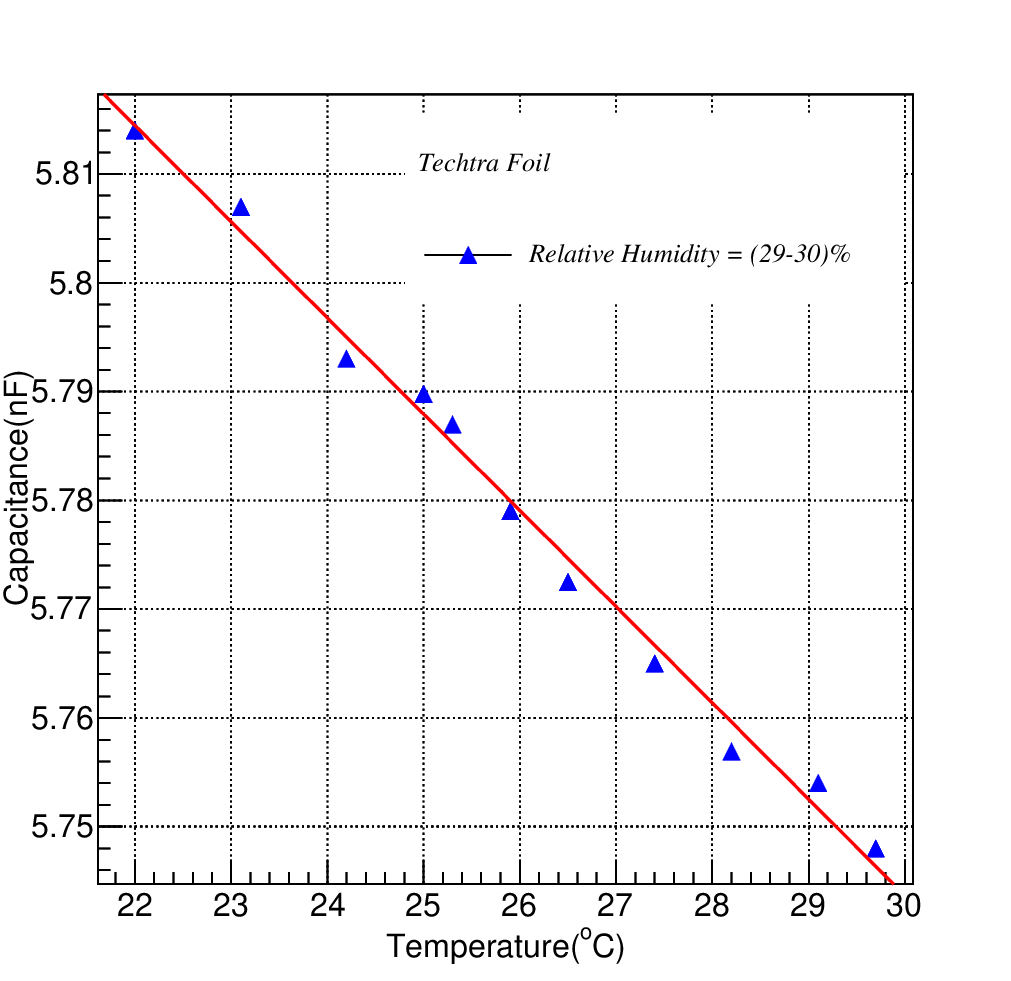}
        \caption{ }
        \label{fig:Temp_Cap}
    \end{subfigure}
    \begin{subfigure}[b]{0.46\textwidth}
        \includegraphics[width=6cm, height=5cm]{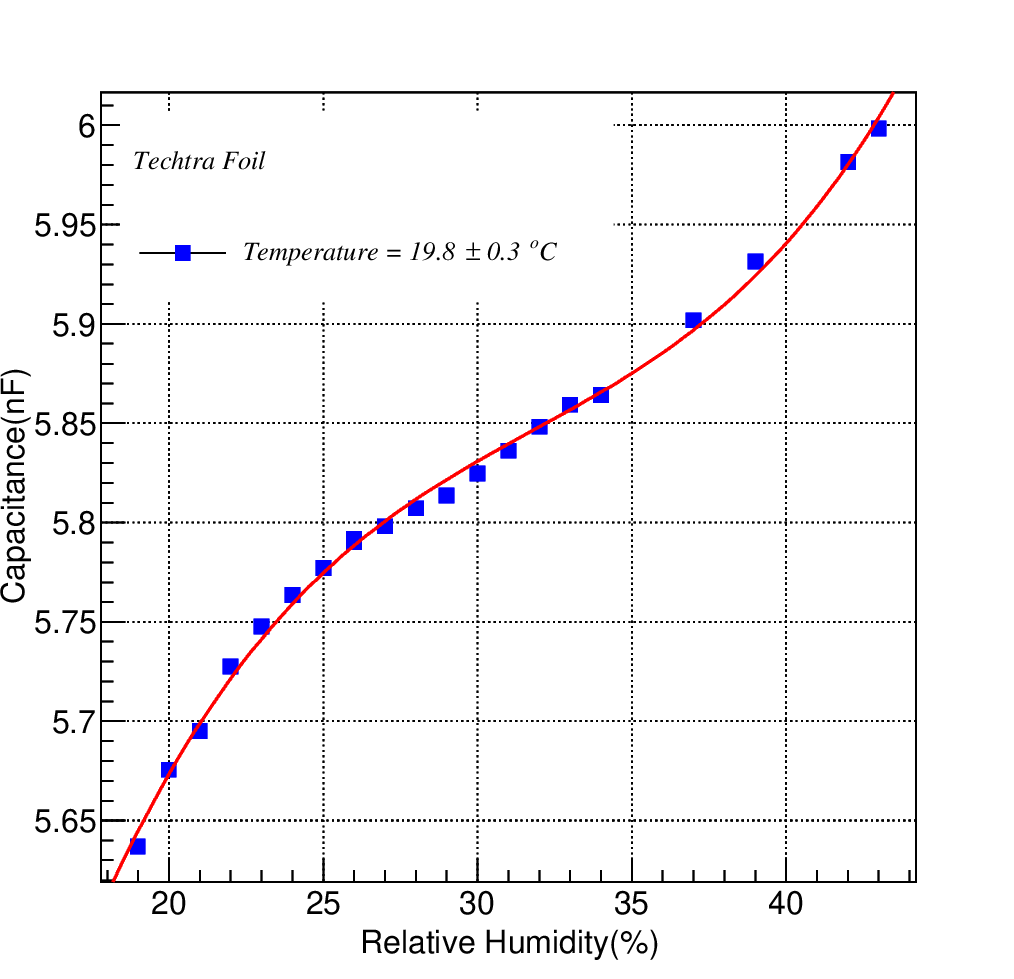}
        \caption{ }
        \label{fig:RH_Cap}
    \end{subfigure}
   \caption{Capacitance variation with: (a) Temperature and fixed RH (29-30)\%, (b) RH and fixed temperature (19.8$\pm$0.3$^o$C). It can be seen that capacitance varies linearly with negative slope for temperature, however  fit to the other graph shows cubic dependancy over RH if other parameters remain fixed.} \label{fig:Temp_RH_Cap}
\end{figure}

Avalanche development and signal production are influenced by the capacitance of GEM foils and the overall GEM detector. As a result, a single foil is examined to see how capacitance reacts to changes in the environment. Temperature, humidity, and capacitance of the foil are measured and three sets of data are obtained. The capacitance of GEM foil varies with humidity and temperature, as seen in Figure \ref{fig:Temp_RH_Cap}. Table \ref{tabel:correlation} demonstrates that capacitance and relative humidity have a strong positive correlation, while capacitance and temperature have a moderate negative correlation.
GEM foil is particularly sensitive to humidity, as the correlation coefficient reveals. This is due to moisture from the gas mixture accumulating on the foil's surface and inside the hole, which increases capacitance and leakage current. Any change in the capacitance of the foils causes a change in the electric field inside holes, which might impact the detector's gain. Temperature, on the other hand, has a modest effect on both capacitance and leakage current. 

\begin{table}[ht]
\begin{center}
\begin{tabular}{ |c|c|c|c| } 
 \hline
  & Temperature & Relative Humidity & Capacitance\\ 
 \hline
 Temperature & 1 & -0.081 & -0.091\\ 
 \hline
 Relative Humidity & -0.081 & 1 & 0.877\\ 
 \hline
  Capacitance & -0.091 & 0.877 & 1\\ 
 \hline

\end{tabular}
\caption{Correlation coeffitient.}
 \label{tabel:correlation}
\end{center}
\end{table}
\section{GEM Detector Performance}
\label{assembly}
\begin{figure}[!ht]
    \centering
        \includegraphics[width=10cm, height=6cm]{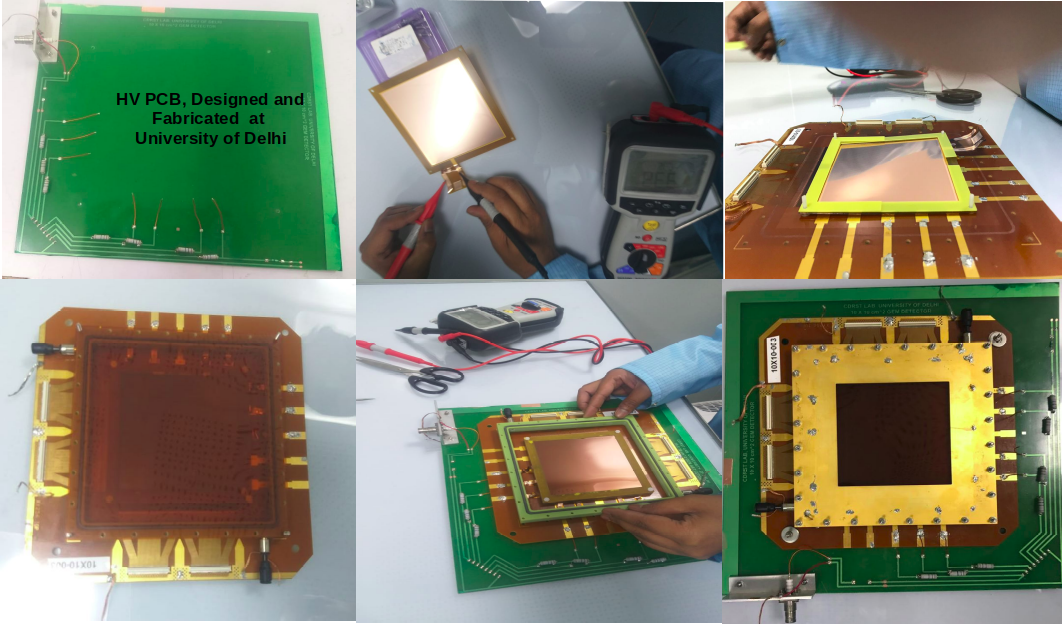}
   \caption{Detector components and assembly procedure for 10 cm $\times$ 10 cm GEM detector.}   \label{fig:assembly}
\end{figure}

The prototype GEM detector with 3/1/2/1 (mm) gas gap [12] is assembled using three fully tested 10 cm $\times$ 10 cm GEM foils, a drift foil (only one side copper plated), and a two-dimensional (2D) readout (RO) board.
The 3 mm gap between the drift foil and the first GEM foil, 1 mm gap between the first and second GEM foils, 2 mm gap between the second and third foils, and 1 mm gap between the third foil and the readout board are all part of this gap arrangement. The detector is built in the same way that the CMS detector was built \cite{six_03}. The assembling technique is shown in Figure \ref{fig:assembly}. 
A voltage divider \cite{gola} network is used to power up the GEM foils and drift foils through a single high voltage (HV) power supply. Each foil is protected using 10 M$\Omega$ resistors against excess current during possible discharges or sparks produced during the operation.
\subsection{Detector characteristics}
\label{detector_charactristics}
The type of gases utilised and the voltage difference across the electrodes have a big impact on how a gas detector behaves. The performance of the detector is also affected by the humidity, temperature, and impurity in the flushing gas. The first test as explained in Section ~\ref{qc4} was performed to see the detector response towards high voltage and fake signal (signal not coming from primary ionization). The gain calibration is explained in Section ~\ref{gain_measurement}, in Section ~\ref{stability} we explained stability while the uniformity of the detector is discussed in Section ~\ref{uniformity}.
\subsubsection{High voltage stress test}
\label{qc4}
\begin{figure}[!ht]
    \centering
    \begin{subfigure}[b]{0.45\textwidth}
        \includegraphics[width=6.5cm, height=5cm]{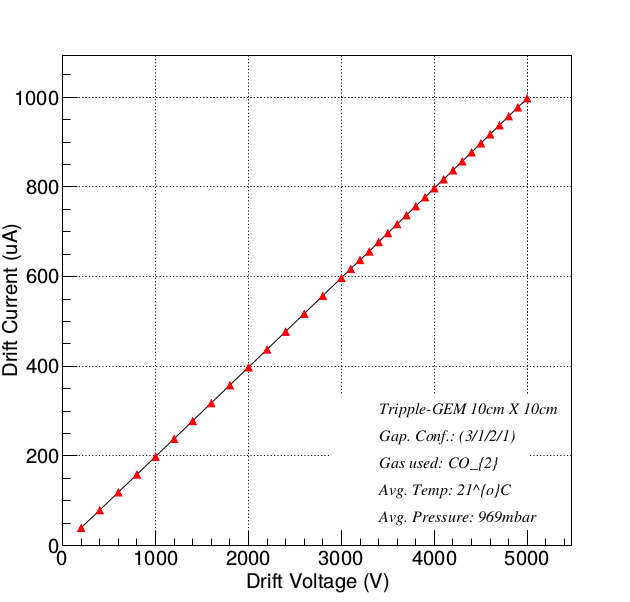}
        \caption{ }
        \label{fig:QC4_HV}
    \end{subfigure}
    \begin{subfigure}[b]{0.45\textwidth}
        \includegraphics[width=6.5cm, height=5cm]{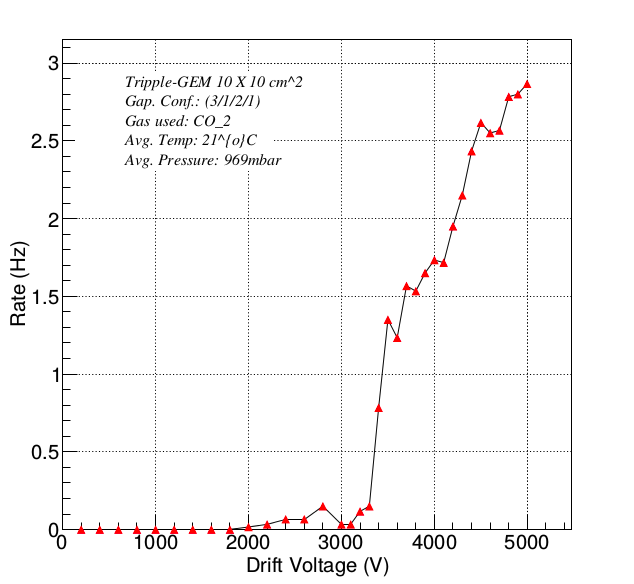}
        \caption{ }
        \label{fig:QC4_rate}
    \end{subfigure}
   \caption{(a) The I-V characteristics of the detector, and (b) shows the variation of spurious signal rate obtained from GEM3 bottom
as a function of the divider voltage} \label{fig:QC4}
\end{figure}

Using a 4.7 M$\Omega$ HV divider and 0.3 M$\Omega$ HV filter, high voltage was applied to the detector to power it. The CMS-GEM community \cite{six_03} uses a 4.7 M$\Omega$ HV divider for a 3/1/2/1 (mm) gap configuration for a triple-layered GEM detector. The detector was flushed with non-amplifying gas, such as CO$_2$, at a flow rate of 3.0 $\ell/h$ before powering up, and the detector's total impedance was measured to be 5.01 M$\Omega$. After minimising electrical noise and stabilising gas flow, the detector was powered up in steps of 100 V to 5000 V using a CAEN-N1471 \cite{N1471h} power supply, and the corresponding divider current was recorded together with a fake signal count from the bottom of the last GEM foil towards readout. The spurious signal rate, which is defined as the rate of signals that do not originate from the ionisation of the gas, is computed from these counts. These spurious signals are caused by fluctuations in the HV power line or the GEM foil surfaces. The detector exhibits ohmic behaviour during operation, as shown in Figure \ref{fig:QC4_HV}. The rate of the spurious signal grows as the voltage across the detector is increased, as shown in Figure \ref{fig:QC4_rate}, with the maximum spurious signal rate measured about 2.8 Hz at a divider current of $\sim$1000 $\mu$A or 5000 V. 

\subsubsection{Gain measurement}
\label{gain_measurement}
The gain of a gaseous detector is defined as the ratio of the final number of electrons produced in the avalanche to the number of primary electrons $n_0$ produced in the drift gap when photon or any charged particle passes through it. This formulation expresses an absolute gain that is difficult to quantify in an actual detector such as the GEM. Indeed, in the case of the Triple-GEM detectors, a number of mechanisms can result in electron loss. Electrons can be collected on copper or Kapton in GEM foils. The so-called transparency \cite{transparency} is a term for these losses. The likelihood that an electron will enter and exit a GEM hole is defined as the transparency of the GEM foil. The degree of transparency is determined by a number of factors, including the hole distribution in the foil, the gas mixture, the voltage across the drift gap, and the voltage across the GEM foils, etc. Because of this, we only assess the effective gain of a Triple-GEM detector, which is defined as $G_{eff}$: 
\begin{equation}
G_{eff}=\frac{n_{eff}}{n_0}=\frac{I_{eff}}{I_0}==\frac{I_{eff}}{e.n_0.R.}
\end{equation}
where $n_{eff}$ ($I_{eff}$) is the number of electrons (effective current) inducing a signal on the anode plane, $e$ is the charge on the electron and $R$ is the rate measured for the incoming particles (photons). To estimate the gain, $n_{eff}$ or $I_{eff}$ must be measured and $n_0$ estimated. The $n_0$ is known since the gain measurement relies on the source used.
We used the AMPTEK Mini-X Silver target \cite{xray} as a source with a well-known energy spectrum. The main emission of the Ag target at 40 kV is 22.8 keV \cite{xray}. The number of primary electrons created in the drift region for Ar-CO$_2$ (70:30) mixture is $\sim$346 while operating X-rays at a voltage of 40 kV and current of 5 $\mu$A which is nicely quoted in \cite{gola}.  The output current ($I_{eff}$) is measured using Keithley Electrometer Model 6517B. For rate (R) measurement detector output is read out using a charge sensitive pre-amplifier (ORTEC 142IH) \cite{ortec142ih} whose output is sent to an amplifier plus shaper (ORTEC 474) \cite{ortec474} unit followed by a discriminator. The resulting digital pulses from the discriminator (CFD-ORTEC 935) \cite{ortec935} are fed to a scaler unit, and the rate plateau is obtained by ramping up the detector slowly using HV source. Before performing the actual measurement, the detector is flushed with Ar-CO$_2$ gas mixtures in a 70:30 ratio at 3.0 $\ell/h$ and left to flush for at least 2 hours. 
\begin{figure}[!ht]
    \centering
    \begin{subfigure}[b]{0.45\textwidth}
        \includegraphics[width=6.0cm, height=5cm]{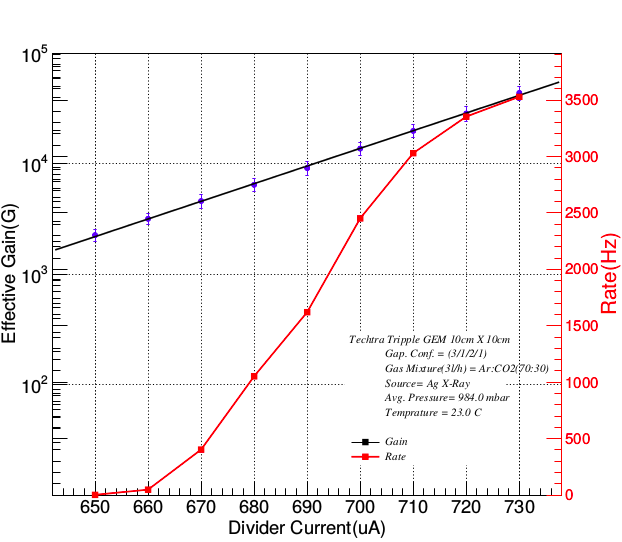}
        \caption{ }
        \label{fig:Gain_7030}
    \end{subfigure}
    \begin{subfigure}[b]{0.45\textwidth}
        \includegraphics[width=6.0cm, height=5cm]{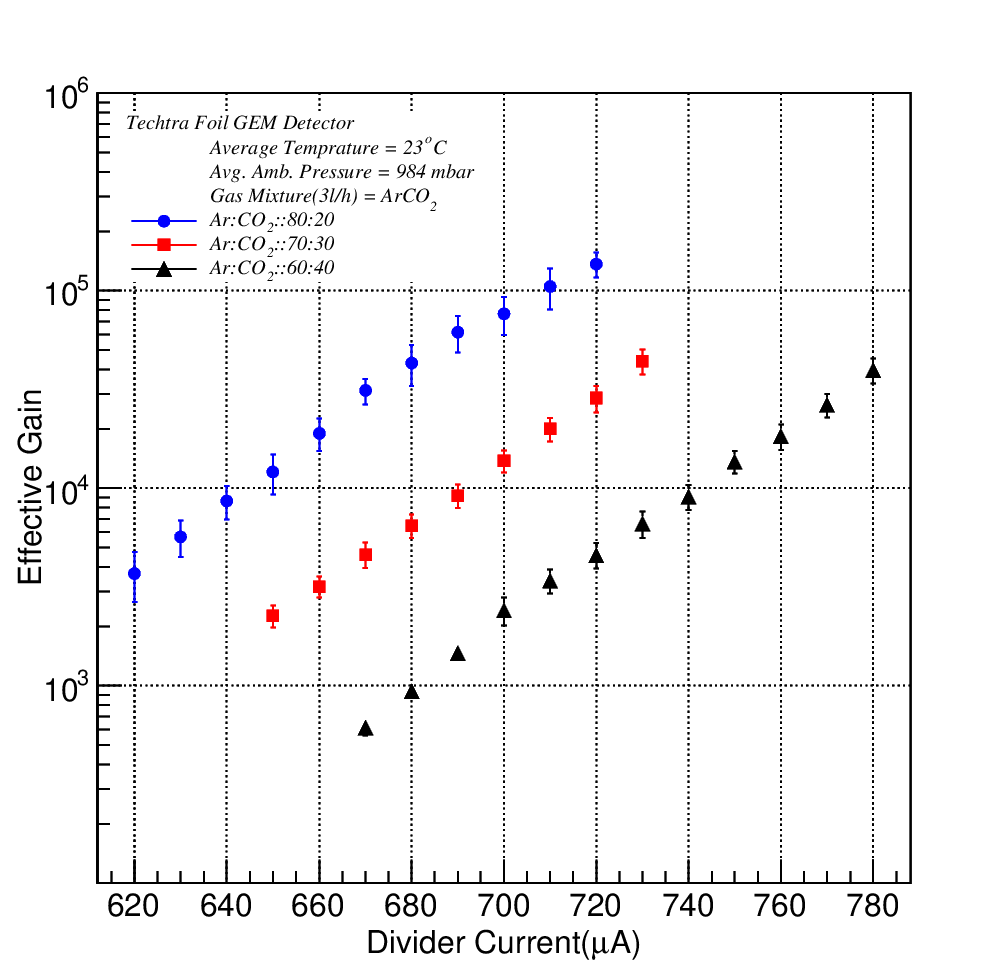}
        \caption{ }
        \label{fig:Gain_3gas}
    \end{subfigure}
   \caption{(a) Shows the gain of the detector with rate, and (b) shows the gain variation with three different Ar and CO$_2$ mixture, as a function of the divider current.} \label{fig:Gain}
\end{figure}
Figure ~\ref{fig:Gain_7030} red curve shows the rate $R$ measured as a function of the divider current at 100 mV of discriminator threshold, while the blue marker along with the error bar shows the effective gain measured on a logarithmic scale as a function of divider current.  Figure \ref{fig:Gain_3gas} shows the comparison plot of effective gain in Ar-CO$_2$ ratio of 80:20, 70:30, and 60:40.

\subsubsection{Gain stability}
\label{stability}

Gain stability is an extremely important parameter for the successful performance of GEM detectors over a long period of time, since time resolution, spatial resolution and efficiency are heavily reliant on the  gain \cite{one}. Even small fluctuations in gain can have a significant impact on these parameters. While keeping the detector at a constant 700 $\mu$A of divider current and X-ray flux constant over the period of time we observed variations in gain and stability over the course of 30 hours. In previous study we have seen the flux has almost no impact on the gain of the detector ~\cite{ashaq_single_mask} ~\cite{asar3030} when it becomes stable after few hours.
\begin{figure}[!ht]
    \centering
        \includegraphics[width=10cm, height=7cm]{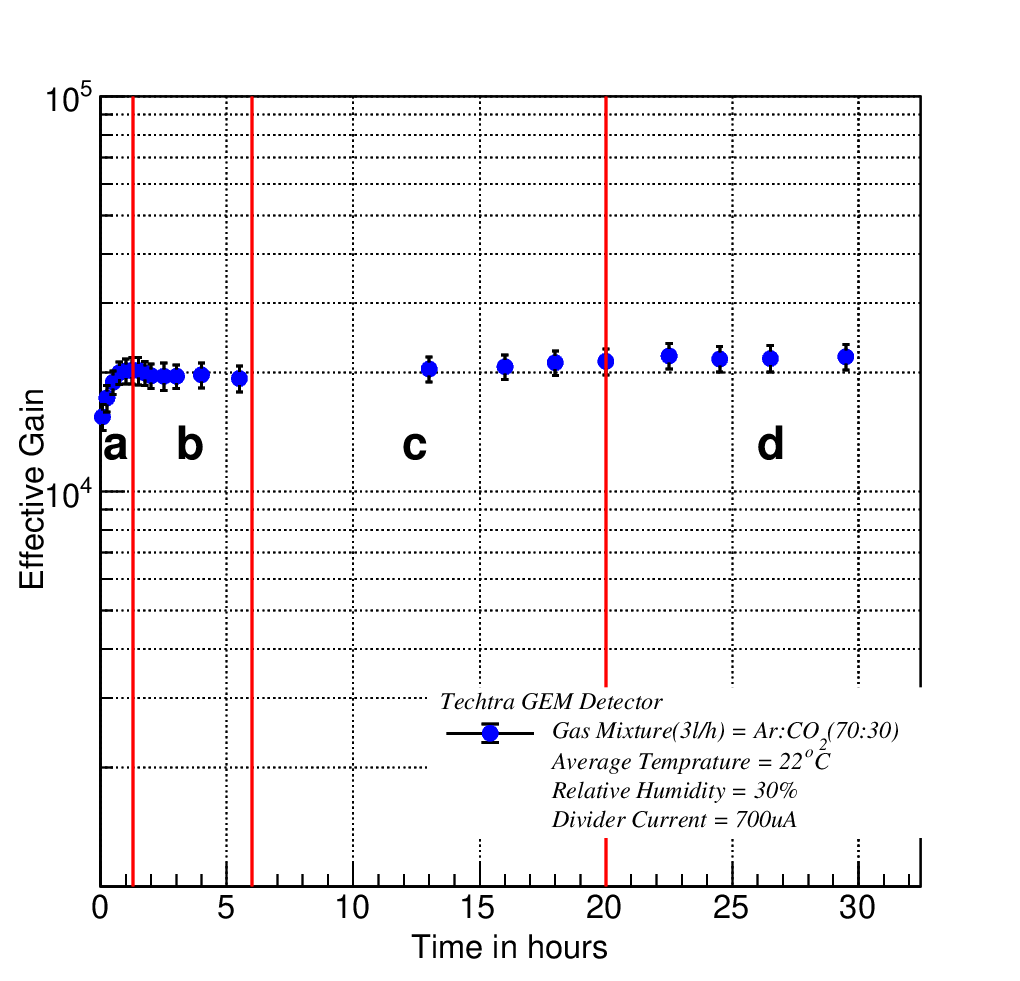}
   \caption{Effective gain stability curve with time. Plot is divided in four regions to compare the change in gain with time where it shows significant rise of 30$\%$ in first one hour.}   \label{fig:stability}
\end{figure}
 Figure \ref{fig:stability} illustrates the changes in gain over time. After an hour, the gain increases by 30$\%$, and in the subsequent 17 hours, it increases by 2.6$\%$ with small variations, and the gain remains stable for the last 12 hours. Variations in gain for the first hour can be linked to charging up phenomena caused by avalanche charge deposition on the GEM foils ~\cite{charging_up}. This charging up is also seen in the foil's leakage current test. Gain during the next 17 hours, on the other hand, is linked to the GEM foil's polarisation phenomenon, which is well understood and can be found elsewhere \cite{gola_charging}.
For constant incoming photon flux, a similar trend in rate(7.5 kHz to 9.4 kHz) and detector current(6.4 nA to 10.5 nA) is seen over first one and half hour of time, leading to the conclusion that the detector is less efficient in the early hours and reaches its maximum and stable plateau after a few hours or complete charging of foils.
\subsubsection{Gain uniformity}
\label{uniformity}
\begin{figure}[!ht]
    \centering
    \begin{subfigure}[b]{0.45\textwidth}
        \includegraphics[width=5.0cm, height=4cm]{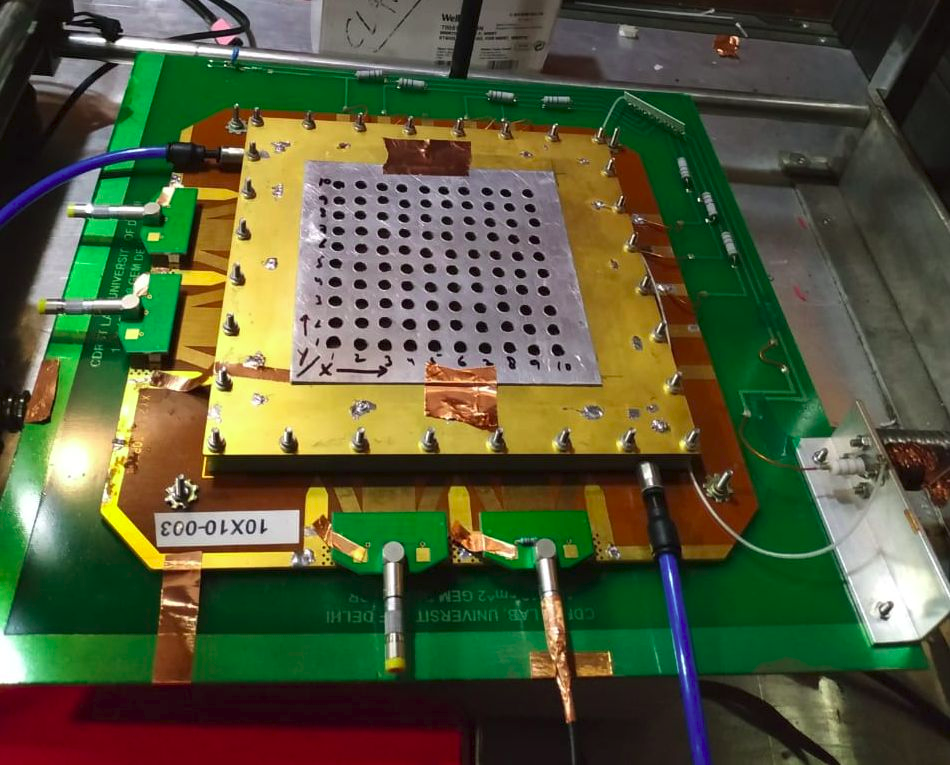}
        \caption{ }
        \label{fig:uniformity_sector}
    \end{subfigure}
    \begin{subfigure}[b]{0.45\textwidth}
        \includegraphics[width=5.0cm, height=4cm]{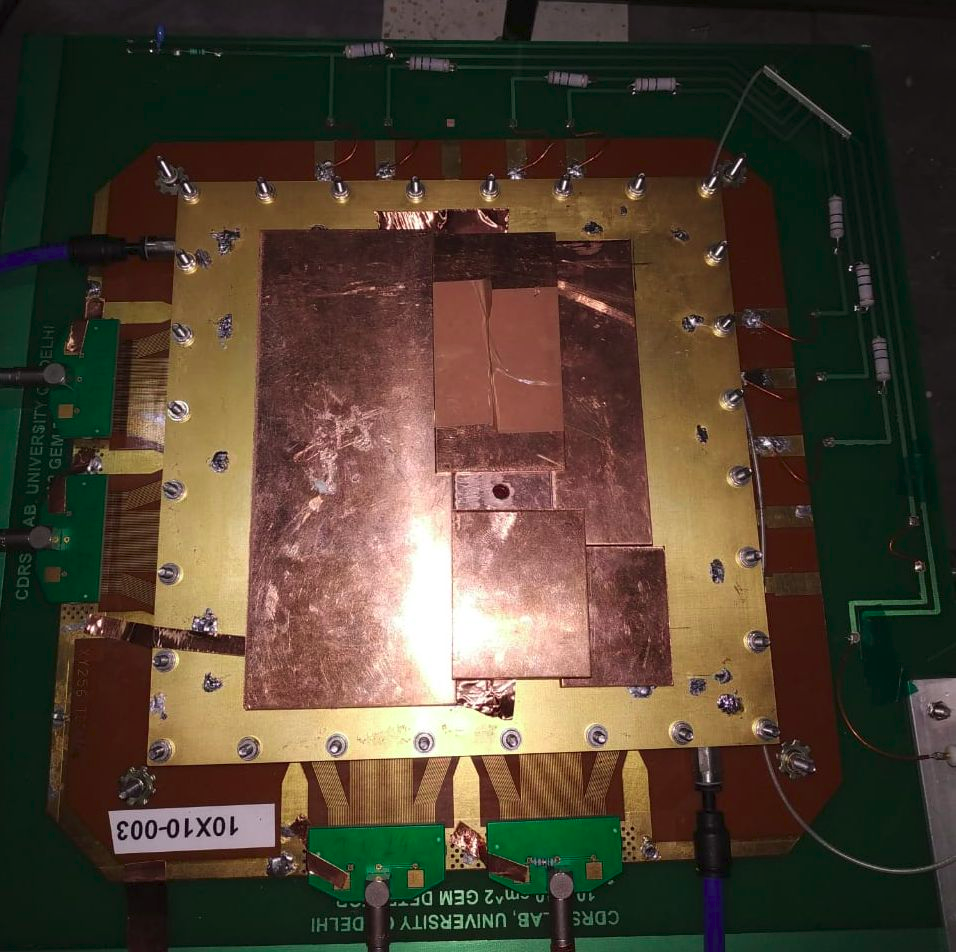}
        \caption{ }
        \label{fig:uniformity_shield}
    \end{subfigure}
   \caption{(a) Shows the the setup for detector uniformity measurements, and (b) shows the single cell exposer using 6mm thick copper shielding.} \label{fig:uniformity}
\end{figure}

\begin{figure}[!ht]
    \centering
    \begin{subfigure}[b]{0.45\textwidth}
        \includegraphics[width=6.0cm, height=5cm]{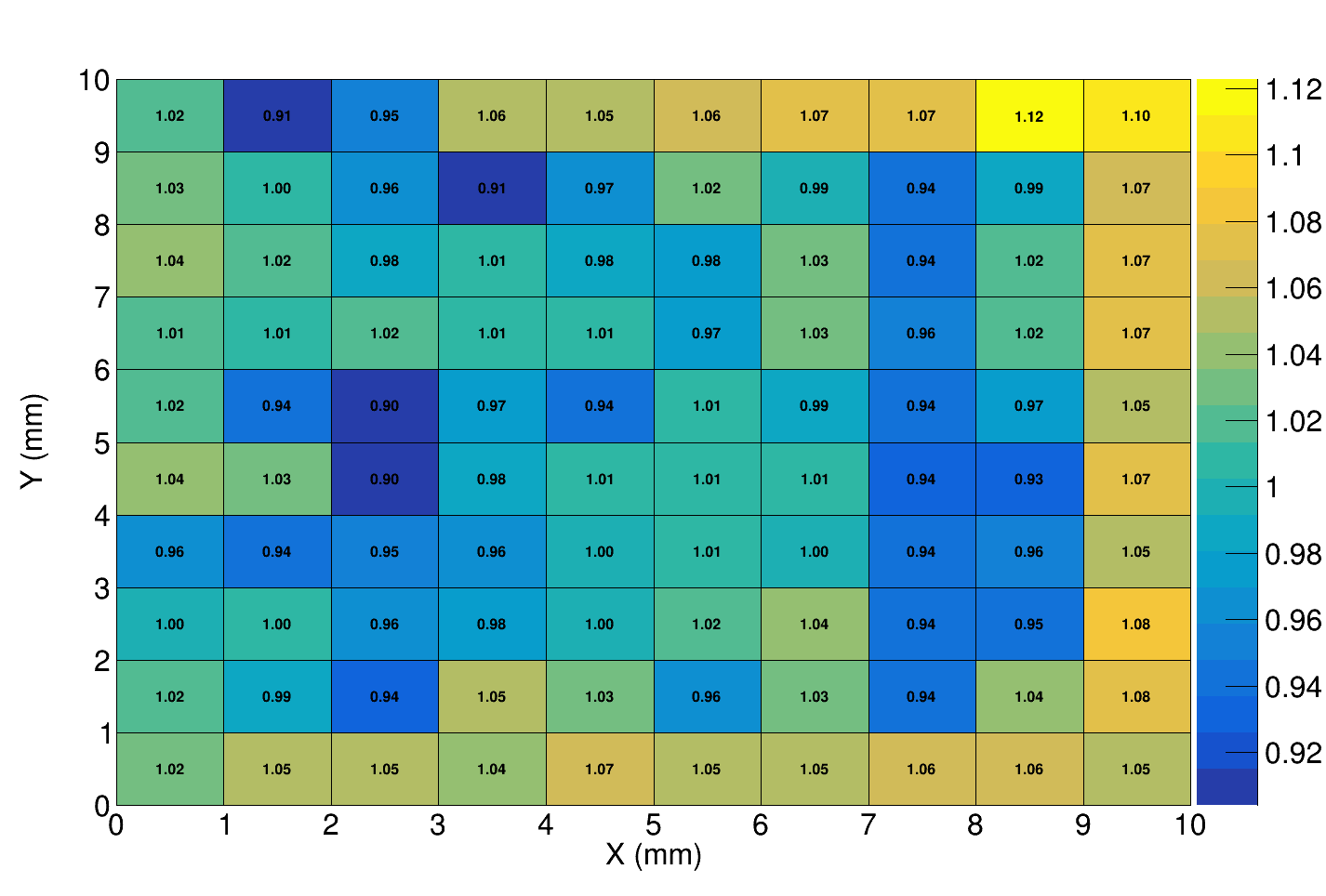}
        \caption{ }
        \label{fig:uniformity_2D}
    \end{subfigure}
    \begin{subfigure}[b]{0.45\textwidth}
        \includegraphics[width=6.0cm, height=5cm]{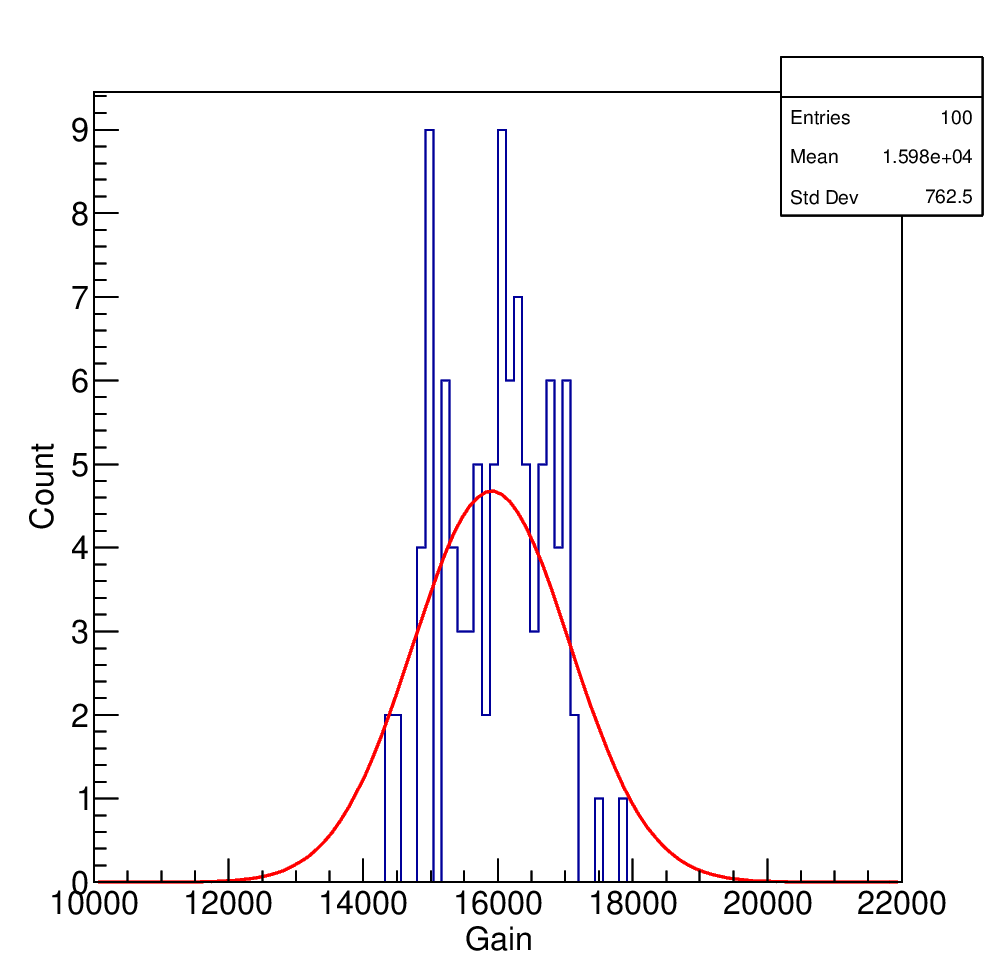}
        \caption{ }
        \label{fig:uniformity_histo}
    \end{subfigure}
   \caption{(a) Gain uniformity map of the detector using 2D histogram, and (b) shows the 1D histogram of 100 cells fitted using Gauss function.} \label{fig:uniformity_histo2}
\end{figure}

The usage of the GEM detector for imaging is one of the significant studies presented in this work. The detector's response must be uniform within an acceptable uncertainty to meet this criteria. A non-uniformity test was carried out in order to determine the detector's response independently. As illustrated in Figure \ref{fig:uniformity_sector}, the detector was split into 100 1 cm $\times$ 1 cm sectors in the XY plane. The X-ray source was then focussed at the centre of each cell using a suitable collimator. To radiate a single sector, a 6 $mm$ thick copper shield with a single hole is utilised to cover the whole detector, as illustrated in Figure \ref{fig:uniformity_shield}, preventing radiation from reaching other areas. 
Figure. \ref{fig:uniformity_2D} illustrates the effective gain map for each sector of the GEM detector at the divider current of 710 $\mu$A. The effective gain is observed to be higher at the corners of the detector while slightly lower in the middle. The histogram shown in Figure. \ref{fig:uniformity_histo2}  is filled with the mean effective gain value from each sector and fitted with Gaussian function. The mean and sigma from the fit is found to be $1.598\times10^4$ and $0.092\times10^4$ respectively. These values are used to calculate the non-uniformity of the detector using $\sigma$/$\mu$ which came out to be $5.76\%$ which is within an acceptable range for our requirements.
The non-uniformity of gain throughout the whole surface of the GEM detectors is due to a number of reasons. This includes variations in gas gaps caused by inaccuracy in stretching, GEM foil quality, non-uniformity or imperfections in holes over the foil region, and so on. Similar behavior of gain uniformity have been observed with the detector using CERN and Micropack GEM foils \cite{gola}. 
\section{Imaging with GEM}
\label{imaging}
The spatial and time resolutions of GEM detectors are well-known.
These characteristics, together with their high flux handling capability, make them ideal candidates for imaging detector applications.
Several attempts to evaluate the imaging capabilities of these detectors have been performed in the past ~\cite{imaging_01}~\cite{pct}. We did a detailed investigation for imaging utilising our built-in GEM detector using Tectra and Micropack foils to better evaluate its imaging capabilities. A completely tested GEM detector with all known parameters was employed for this purpose. As illustrated in Figure 8, the 2D readout board contains 256 strips with a 0.39 mm pitch along the X and Y axes. To readout the signals, the RO board is attached to four Panasonic male connectors with 128 channels. To gather the charge from the readout at 6k samples/s, a 256 channel Techtra GEM board ~\cite{gemboard} is utilised, with two successive strips packaged together to make it compatible with the readout circuits. Four DDC24 ~\cite{ddca} 20 bit 64 channels current input analogue to digital converter (ADC) make up the GEM readout electronics board. It has a Xilinx Spartan-3 FPGA for controlling the numerous components, as well as a communication module that allows it to connect with the computer at 100 Mbps. 
\begin{figure}[ht]
    \centering
    \begin{subfigure}[b]{0.45\textwidth}
        \includegraphics[width=6.0cm, height=5cm]{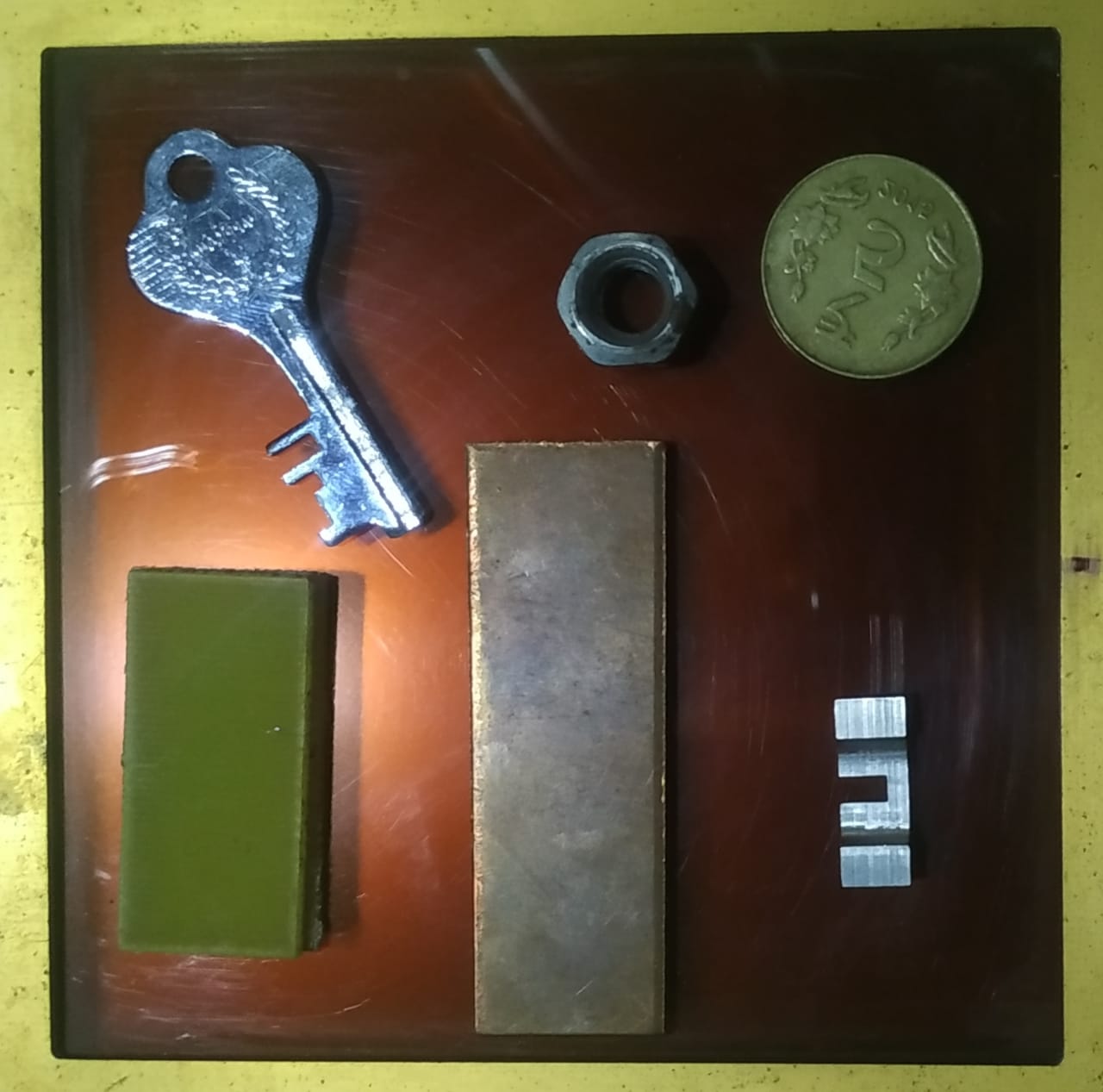}
        \caption{ }
        \label{fig:material}
    \end{subfigure}
    \begin{subfigure}[b]{0.45\textwidth}
        \includegraphics[width=6.0cm, height=5cm]{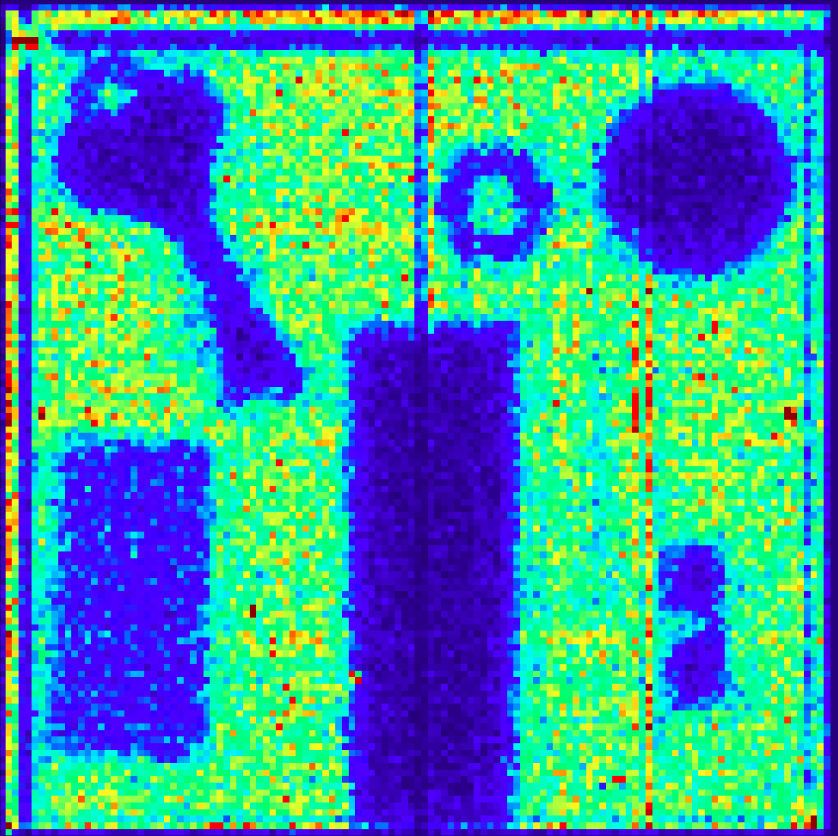}
        \caption{ }
        \label{fig:material_image}
    \end{subfigure}
   \caption{(a) Objects of different materials under scan and (b) reconstructed image of the scanned objects.} \label{fig:uniformity_histo2}
\end{figure}

The detector was flushed with Ar-CO2 (70:30) gas mixture at 2$\ell/h$, and powered up to set the gain of $\sim$10k.  X-ray source was tuned to operate at 5.0 kHz. The detector hardware is provided with dedicated software for detector control and data acquisition. The amplifier settings such as integration time were set to 200 $\mu$s and charge range up to 1.25 pC.
The data is recorded in raw format, which means that picture reconstruction will require additional data processing. To improve image quality, basic cuts were made across the energy spectrum and over time. We were able to reconstruct the image as shown in Figure ~\ref{fig:material_image} without any offline modifications for various objects such as copper, steel, alloys, and FR4 \cite{fr4} as displayed in Figure ~\ref{fig:material}. The gain of the detector and the number of events gathered have a significant impact on the reconstructed image quality. 
With 20 thousand events or 10.24 million subevents\footnote{Here one subevents corresponds to one hit on each channel of readout:\\ 20000 events X 512 channels = 10240000 subevents}, Figure ~\ref{fig:material_image} has been reconstructed. 
The limited data is due to the problem in the X-ray, which heats up after few minutes of operation and needs to be switched off. This limitations will be sorted out with a better X-ray machine. The other limitation in data taking is coming from the slow readout electronics. This can also be easily taken care of by utilizing a fast readout system in future attempts.
\subsection{Object Identification \& Measurement}
\begin{figure}[ht]
    \centering

    \begin{subfigure}[b]{0.45\textwidth}
        \includegraphics[width=6cm, height=5cm]{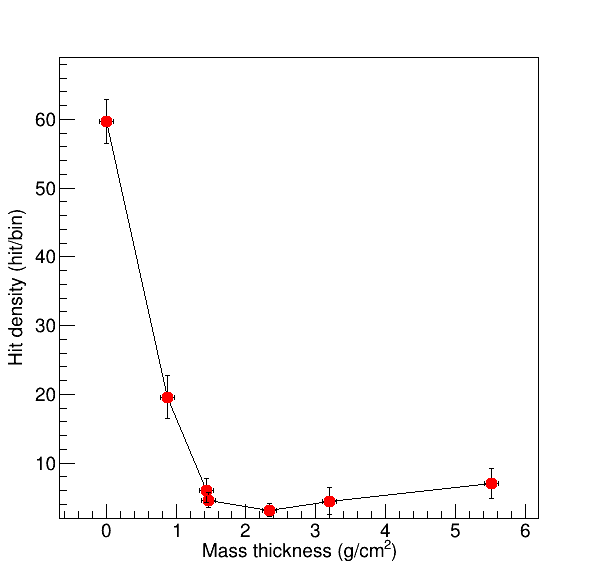}
        \caption{ }
        \label{fig:hit_density}
    \end{subfigure}
        \begin{subfigure}[b]{0.45\textwidth}
        \includegraphics[width=6cm, height=5cm]{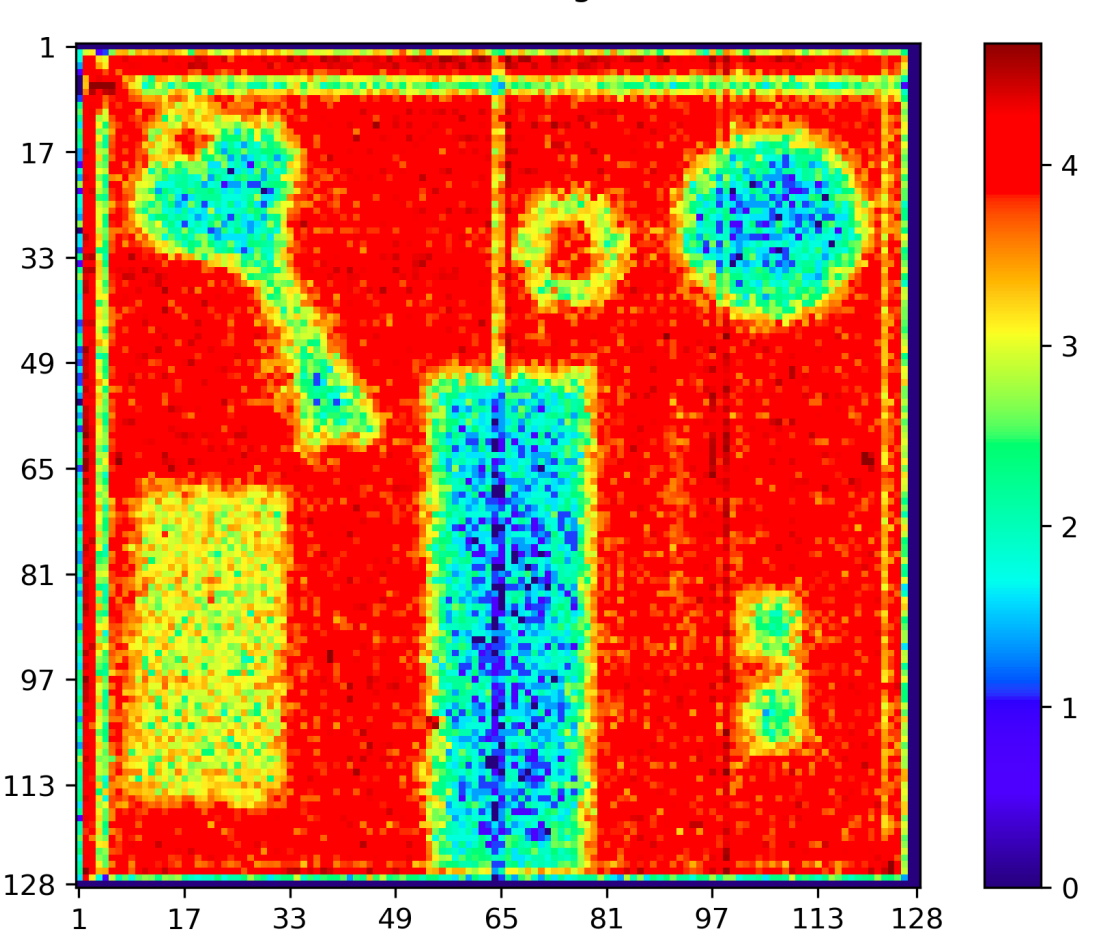}
        \caption{ }
        \label{fig:image_log}
    \end{subfigure}
   \caption{(a) Hit density vs mass thickness. (b) Reconstructed image on logarithmic scale. } \label{fig:uniformity_histo2}
\end{figure}

Materials can be easily identified based on mass attenuation coefficient ($\mu/\rho$) which fits in the formula $I/I_o=e^{-(\frac{\mu}{\rho}x)}$ \cite{nist}, where $x$ is the mass thickness and $\rho$ is the density of the material. $I/I_o$ have been calculated from the reconstructed image of each object for 22.8 keV of X-ray. The last column in Table. \ref{tab:table1} shows the measured values of the hit density\footnote{The number of hits per bin, this has been calculated by taking the sufficient number of bins below objects and averaging over it.} information ($I$) underneath each object. This hit is created by penetrating photons that reaches the detector volume. Figure \ref{fig:hit_density} shows the $I/I_o$ verses mass thickness of the material.  The result displays the expected trend and classifies the difference between metallic and nonmetallic objects clearly. The object having larger thickness throws up larger uncertainty in hit information. This is due to larger scattering (coherent or Compton scattering)\cite{Mansour} probabilities which leads to large statistical fluctuations. Due to small size, sharp curvature, and larger thickness along with multiple scattering issue, the hit reconstruction with very fewer events is poor. This   affects the hit density in case of Nut and Pullout which is visible in Figure \ref{fig:hit_density}.

\begin{table}[ht]
\begin{tabular}{ |p{2.4cm}||p{1.5cm}|p{1.3cm}|p{1.8cm}|p{2.1cm}|p{1.3cm}|  }
 \hline
 Material & Density ($gcm^{-3}$) & Thick. (cm) & Mass Thickness ($g/cm^{2}$ & Hit density $I$ $(hit/bin)$ & $I/{I_o}$\\
 \hline
 \hline
 Kapton  ($I_o$)& 1.42 & 0.005 & 0.007& 59.70 & 1.000\\
 FR4  & 1.84 & 0.48 & 0.88 & 19.59 & 0.328\\
 Key (316)& 8.00 &0.18 & 1.44 & 6.03 & 0.101\\
 Coin (FSS) & 7.70 &0.19 & 1.46 & 4.57 & 0.077\\
 Copper   &8.96  &0.26 & 2.33 & 3.16 & 0.053\\
 Pullout (316) & 8.00 &0.40 & 3.20 & 4.45 & 0.075\\
 Nut (Iron)& 7.87  & 0.70 & 5.51 & 7.10 & 0.119\\
 \hline
\end{tabular}
\caption{Experimental intensity ratio table for materials used.}
\label{tab:table1}
\end{table}
The reconstructed image was also casted on the density scale from least to maximum density. The color of the objects provides a fair distinction between the objects of varying densities in Figure \ref{fig:image_log}.
The estimation of the size of objects is obtained by measuring the object dimension in the number of bins
 where hit density is low. Bin size is calculated by dividing the size of RO pad (100 mm) by total number of readout channels along one axis (128), as Bin size $ = 100/128 = 0.78125$ mm. This factor of 0.78125 is used to calculate the dimension of each reconstructed image. The table \ref{tab:size} shows the size of the reconstructed images obtained from the data and their comparison with the size measured with an electronic vernier caliber. The change in the length or diameter between the actual dimension and the one obtained from imaging is provided in the last column. A direct correlation between the mass thickness of the object and uncertainty in its measured dimensions have been observed. The objects having larger mass thickness have a larger uncertainty as expected. This is due to the larger scattering (coherent or Compton scattering) probabilities from objects having a higher mass thickness and lateral surface exposure. 
\begin{figure}[ht]
    \centering
        \includegraphics[width=5cm, height=6cm, angle =-90]{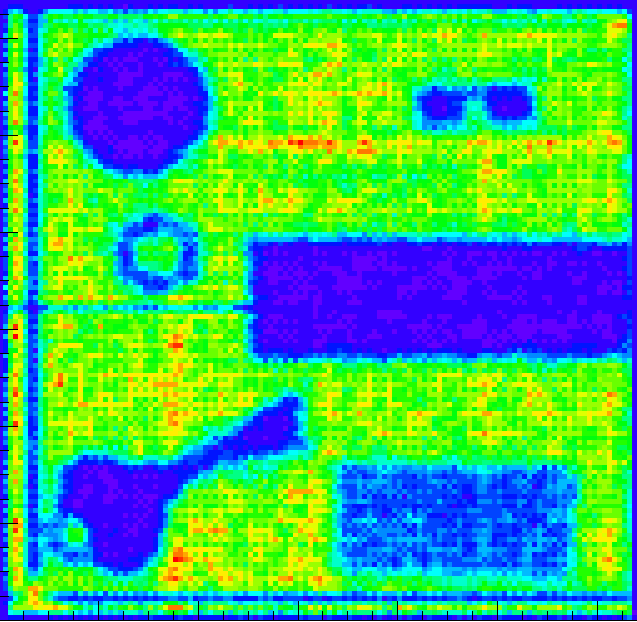}
   \caption{Image after bin fitting and corrections made.}   \label{fig:image_fit}
\end{figure}
Since these known effects can be easily fixed offline, therefore, an algorithm has been developed to take care of this. In this algorithm a bin wise fitting was performed to correct the sharp edge boundaries which leads to the enhancement of the quality of the image. The final image obtained after correction is shown in Figure \ref{fig:image_fit} and the corresponding measurement of the dimensions of the image/objects is shown in Table \ref{tab:size}. The detector constructed with Micropack or Techtra foils gives similar image reconstruction and dimension measurements. 

\begin{table}[ht]
\begin{centering}
\begin{tabular}{ |p{3.1cm}||p{1.2cm}|p{1.2cm}|p{1.2cm}|p{1.2cm}|p{1.2cm}|p{1.2cm}|}
 \hline
\multirow{2}{*}{Material} & \multirow{2}{*}{A.D}  & \multirow{2}{*}{M.T} & \multicolumn{2}{|c|}{Meaured Dimension} & \multicolumn{2}{|c|}{Change ($\%$) } \\
\cline{4-7}
& & & B.C & A.C & B.C & A.C\\

\cline{4-7}
 \hline
 \hline
  FR4 (L)  & 38.63    & 0.88 & 38.81 & 38.28 & 1.42 & 0.91\\
 FR4 (W) & 20.49    &  & 20.32 & 20.31 & 0.83 & 0.88\\
 \hline
 Key (L)& 47.50   & 1.44 & 47.97 & 47.47 & 0.99 & 0.06\\
 Key (W)& 22.04   &  & 22.31 & 22.15 & 1.23 & 0.49\\
 \hline
 Coin (dia.) & 23.05   & 1.46 & 23.45 & 23.44 & 1.79 & 1.69\\
 \hline
 Copper (L)   & 61.41 & 2.33 & 61.53 & 61.49 & 0.20 & 0.13 \\
 Copper (W)  & 20.1 &    & 20.61 & 20.09 & 2.53 & 0.05\\
 \hline
 Pullout (L) & 19.03   & 3.20 & 19.49 & 19.14 & 2.42 & 0.58\\
 Pullout (W) & 7.15   &  & 7.41 & 7.23 & 3.63 & 1.11\\
 \hline
 Nut (Outer dia.)& 13.16      & 5.51 & 14.82 & 13.28  & 12.61 & 0.91\\
 Nut (Inner dia.)& 6.50      &  & 4.80 & 6.25  & 26.15 & 3.85\\
 \hline

\end{tabular}\\
\caption{Objects size measurements and percentage change from actual values.Where Abbreviations used are, A.D: actual dimension in mm, M.T: mass thickness in $g/cm^{2}$, B.C: before corrections, A.C: after corrections.}
\label{tab:size}
\end{centering}
\end{table}

\section{Conclusions}
\label{conclusion}
GEM foils of 10 cm$\times$10 cm have been tested both optically and electrically to use them in our future experiments for imaging. The optical test reveals the uniformity in the hole pattern and found no significant defects. The mean hole diameter measured to be 73.98 $\mu$m (top) and 65.42 $\mu$m (bottom) in copper while 53.37 $\mu$m (top) and 50.78 $\mu$m (bottom) in Kapton. This asymmetry in hole diameter is due to the single mask etching process. The leakage current is less than 1 nA for all the foils at the applied voltage of 600V at 21$^o$C, and RH of 30$\%$. In the longer duration test, the leakage current have been found to decrease exponentially for the first hour and becomes stable afterwards.  The leakage current is highly dependent upon ambient conditions; it increases linearly with temperature and exponentially with the relative humidity. The HV stress test of the GEM detector shows a good performance with a very low spurious rate. The measured effective gain and stability study shows the expected behavior. The gain uniformity of the detector is quite good with only 5.7$\%$ non-uniformity across the total area of the detector. An effort has been made to image various objects with the GEM detector. The initial results from the imaging is encouraging. All the objects of varying densities can be clearly distinguished with contrast. The dimensions of reconstructed images shows a good agreement with actual dimensions within 1$\%$ of variation for most of the objects. The studies performed in this paper is a proof of principle of successfully using GEMs for imaging objects of varying densities. In the next iteration of the study, we plan to improve the X-ray setup, the readout system, and the reconstruction algorithm to achieve a sub-millimeter resolution of the image.
\section*{Acknowledgements}
We would like to acknowledge the funding agency, Department of Science and Technology (DST), New Delhi for providing generous financial support. We would also like to thank the Delhi University - Institution of Eminence (grant no. IoE/FRP/PCMS/2020/27) to support this work.\\ Asar Ahmed would like to express gratitude to University Grant Commission for providing fellowship during the course of Ph.D. work, including this one.

\end{document}